%% file: main_IML_SP_arxiv.tex
\documentclass[journal]{IEEEtran}

\usepackage{cite,graphicx,setspace}

\usepackage{amsmath}
\usepackage{dsfont}
\usepackage{amsthm}
\usepackage{color,subfigure,algorithm,algpseudocode}

\DeclareMathOperator*{\argmax}{arg\,max}
\DeclareMathOperator*{\argmin}{arg\,min}

\newtheorem{theorem}{Theorem}
\newtheorem{lemma}{Lemma}

\input{defns.tex}

\input{notations.tex}

\begin{document}
%
\title{On convergence  and optimality  of maximum-likelihood APA}

\author{Shirin Jalali \thanks{SJ is with Electrical and Computer Engineering Department, Rutgers University. E-mail: shirin.jalali@rutgers.edu} \and 
Carl Nuzman \thanks{CN is with Nokia Bell Labs, Murray Hills, NJ. E-mail: carl.nuzman@nokia-bell-labs.com} \and
Yue Sun \thanks{YS worked on this project while he was with Nokia Bell Labs, Murray Hills, NJ, USA. E-mail: yuesun9308@gmail.com}}
\maketitle

\begin{abstract}

Affine projection algorithm (APA)  is a well-known  algorithm in adaptive filtering applications such as audio echo cancellation.  APA relies on three parameters: $P$ (projection order), $\mu$ (step size) and $\delta$ (regularization parameter).  It is known that running APA for a fixed set of parameters  leads to a tradeoff between    convergence speed and accuracy. Therefore, various methods for adaptively setting the  parameters have been proposed in the literature.  Inspired by  maximum likelihood (ML) estimation,  we derive a new ML-based approach for adaptively setting the parameters of APA, which we refer to as ML-APA. For memoryless Gaussian inputs, we fully characterize the expected misalignment error  of ML-APA as a function of iteration number and show that it converges to zero as $O({1\over t})$. We further prove that the achieved error is asymptotically optimal.    ML-APA updates its estimate once every $P$ samples. We also propose incremental ML-APA (IML-APA), which updates the coefficients at every time step and outperforms ML-APA in our simulations results.    Our simulation results verify the analytical conclusions for memoryless inputs and  show that the new algorithms also perform well for strongly correlated input signals.
\end{abstract}


%
\IEEEpeerreviewmaketitle


\section{Introduction}

Adaptive filtering is a fundamental problem  in signal processing with many applications. In a generic system identification scenario, the response of an unknown linear system to a known input signal is used to estimate the impulse response of the unknown system. 
During the past sixty years, there has been extensive research on developing efficient high-performance adaptive filters. The work has led to the development of various well-known algorithms, such as least mean squares (LMS), normalized least mean squares (NLMS), recursive least squares (RLS) and the affine projection algorithm (APA).  Among these methods, the APA algorithm has been found to be the most efficient method that is widely used in practice as well, because on one hand it is computationally efficient and on the other hand is able to offer good performance in terms of convergence speed and accuracy. 

To run the APA algorithm one needs to specify three variables: the projection order ($P$), the regularization parameter ($\delta$) and the step size ($\mu$). It is well-understood in the literature that statically  setting these parameters, especially $\mu$ and $\delta$,  leads to  a trade-off between convergence speed and accuracy of the algorithm. Therefore, to achieve a good performance, one needs to set these parameters adaptively as the algorithm proceeds. This has led to a host of different solutions that are developed for adaptively setting the aforementioned parameters. (Refer to Section \ref{sec:related-work} for a brief overview of such solutions.)

\begin{figure}[h]
\centering
        \includegraphics[width=0.4\textwidth]{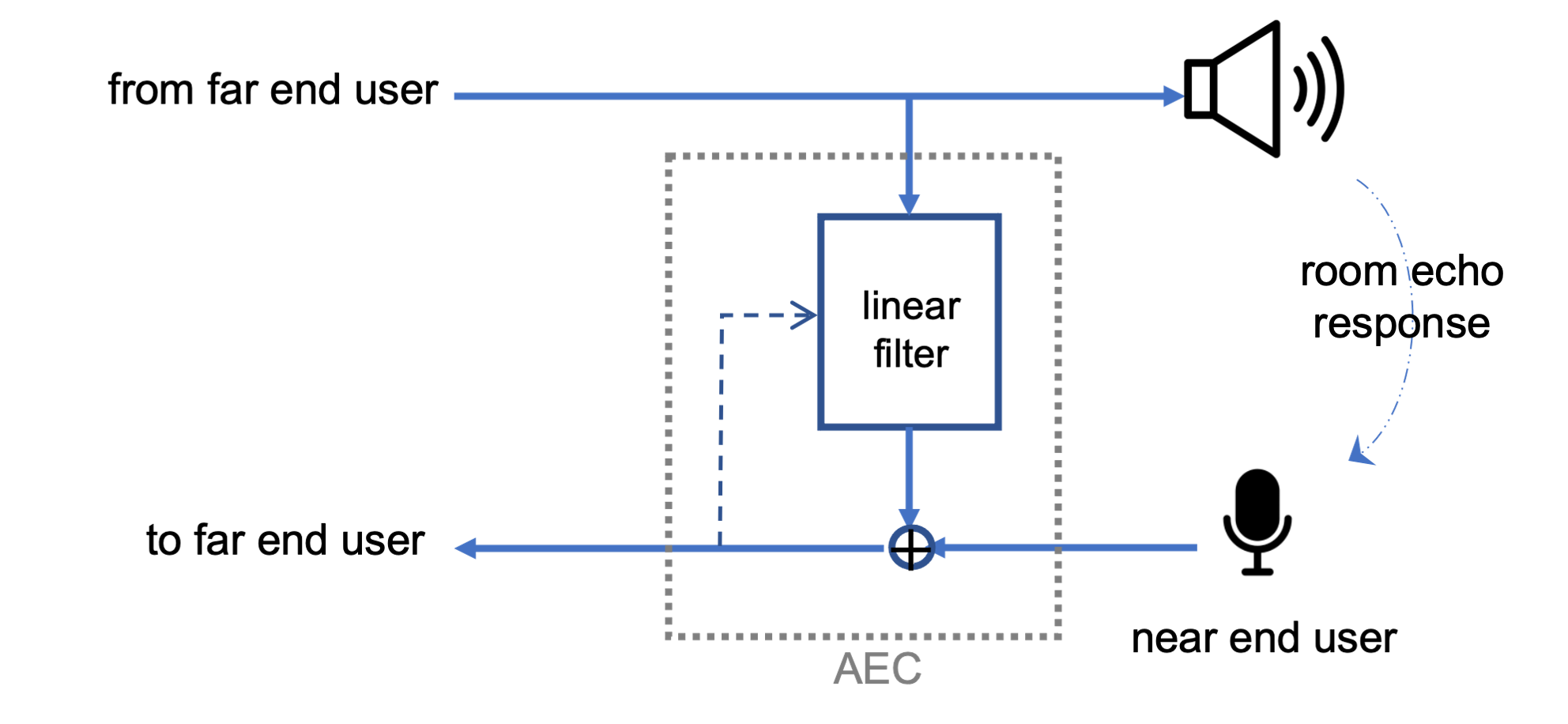}
         \caption{Audio echo cancellation in the presence of multiple speakers and multiple microphones}
         \label{fig:AEC}
\end{figure}

In this paper, inspired by the maximum likelihood (ML) estimation framework,  we present ML-APA, a novel approach inspired by maximum likelihood (ML) estimation, that has provable optimality properties. In this approach, $P$ is fixed, $\mu$ is set to unity, and only $\delta$ is adapted. The ML framework suggests a specific formula for controlling $\delta$ that has close connections with other optimized APA methods in the literature.  The key contribution of this paper is the following: We fully and rigorously characterize the  convergence behavior of ML-APA as a function of $t$ (the iteration number) and show that it asymptotically matches with that of ordinary least squares (OLS) solution, which is the  non-adaptive ML solution. This confirms the asymptotic optimality of ML-APA and shows that for the specified choice of parameters it is able to recover the parameters asymptotically at an optimal rate.  ML-APA is developed for the periodic APA setup, where the coefficients are updated once every $P$ samples. We also derive an alternative version ML-APA, which we refer to as incremental ML-APA (or IML-APA), which updates the coefficients at each iteration.  We  show experimental evidence that both ML-APA and IML-APA have excellent performance in acoustic echo cancellation, an adaptive filtering application where the input signal can be modeled as having  significant memory.  (Ref to Fig.~\ref{fig:AEC} for a block diagram of a  two-way audio communication systems with EC.)

\subsection{Problem statement}
Let $x_t$ and $y_t$ denote the transmitted signal and received signal at time $t$, respectively. Assume that $x_t$ and $y_t$ are related through a linear time-invariant filter modeled by $\wv^*\in\mathds{R}^{L}$, such that  $y_t$ is a noisy observation of a  linear function of the input vector $\xv_t= [x_t,\ldots,x_{t-L+1}]\in{R}^{L}$ as
\begin{equation}
y_t=\xv_t^T\wv^*+z_t, \label{eq:lti_system}
\end{equation}
where $(z_t)_t\stackrel{\rm i.i.d.}{\sim} \Nc(0,\sigma_z^2)$ denotes the independent  noise.   The  goal of an adaptive filter is to adaptively estimate $\wv^*$ from measurements $(\xv_{t'},y_{t'})_{t'=1}^t$.  Assuming that the filter $\wv^*$ is static and does not change, ideally,  at time $t$,  the adaptive filter's goal is  to minimize the following cost function:
\begin{align*}
\ell_t(\wv)=\sum_{i=1}^t(\xv_i^T\wv-y_i)^2.
\end{align*}
However, solving this optimization at every iteration is computationally infeasible. Moreover, in many application including AEC, it is not reasonable to assume that the filter coefficients stay constant.   Therefore, in designing adaptive filters, the goal is to find an algorithm that \emph{adaptively} estimates $\wv^*$, such that at every iteration it updates its current estimate by only using the latest $P$ observations, where $P$ is a small number (typically smaller than $10$).  At time $t$, the adaptive filter receives as input: i) $\xv_t$: input vector at time $t$ (e.g. the loudspeaker signal vector in AEC), ii)  $y_t$:  the output value at time $t$ (e.g., the microphone signal in AEC), iii) $(\xv_{t'},y_{t'})$, $t'=t-1,\ldots,t-P+1$ and iv) $\wv_t$, and outputs an updated estimate of the filter coefficients $\wv_{t+1}$.  The performance of such a adaptive filter can be evaluated in terms of its i) convergence rate, i.e., how fast it can learn the filter coefficients,  ii) accuracy, i.e., how well it can learn the coefficients in steady state (i.e., after enough time has elapsed) and iii) computational complexity and memory requirements.  The ultimate goal in designing adaptive filters is to have a computationally-efficient fast-converging method that is also asymptotically accurate ($\|\wv_t-\wv^*\|_2\to 0$, as $t\to\infty$). 

\subsection{Related work and our contributions}\label{sec:related-work}

 Numerous studies have investigated adaptive tuning of the parameters of  APA and analyzing its convergence, often in the context of acoustic echo cancellation (AEC). Many approaches seek to choose parameters that greedily minimize expected misalignment error in each iteration, using various models and approximations \cite{myllyla2002pseudo,shin2004variable,rey2004analysis,rey2006optimum,rey2007variable,yin2010variable}. In the case of $P=1$, APA reduces to NLMS, which has its own extensive literature on parameter adaptation \cite{benesty2006nonparametric,paleologu2008variable,huang2011new,paleologu2015overview, ciochinua2016optimized}. Adaptive adjustment of projection order $P$ is considered in \cite{kim2009affine,gonzalez2012affine}.  The variable step-size approach (VSS-APA) \cite{paleologu2008variable,albu2011variable} takes input variability and acoustic double talk into account, while \cite{lee2012optimal} and \cite{lee2012scheduled} optimize step-size based on MSE analysis of misalignment error. 
Our approach is mostly closely related to JO-APA \cite{ciochinua2015optimized}, which jointly optimizes the step-size and regularization parameters to greedily minimize expected misalignment, resulting in unit step-size and an adaptive regularization equal to the inverse of a misalignment to noise ratio (MNR). JO-APA includes a method for estimating this ratio, and shows excellent performance in some empirical examples including the case of step changes to filter impulse response.

Building on JO-APA, our distinct contribution is to develop  an adaptive APA framework (ML-APA) whose convergence behavior can be explicitly characterized, and shown to be asymptotically optimal, under idealized condition.  In particular, our analysis holds when i) input signals are i.i.d.,  ii) the true MNR is provided by an oracle, and iii) block updates are performed every $P$ samples. Our analysis shows there is only a  negligible asymptotic gap between ML-APA and non-adaptive offline least squares (OLS) minimization, confirming that ML-APA converges to the true channel response at an optimal rate. 

To validate the efficacy of ML-APA in practice, we conducted empirical experiments focused on acoustic echo cancellation, where the input signal, taken from a human speech database, exhibits strong memory. Our tests highlight ML-APA's exceptional performance in echo cancellation tasks, even though the inputs are far from  i.i.d., and the MNR is estimated rather than oracle-provided. We also propose and demonstrate the efficacy of a related algorithm, IML-APA, that updates the filter coefficient after each new observation

\subsection{Notation}

Throughput the paper bold letters indicate vectors. Matrices are denoted by upper-case letters, such as $A$ and $X$. For  an $m \times n$ matrix $A$, let  $\sigma_{\min} (A) $ and $\sigma_{\max}(A)$ denote the minimum and maximum singular values of $A$. Sets are denoted by calligraphic letters, such as $\Xc$ and $\Yc$. 

\subsection{Organization of this paper}
In Section \ref{sec:apa}, we analyze the convergence behavior of regularized APA. Inspired by the analysis, in Section \ref{sec:ML-alg}, we propose the  ML-APA method, which is an online MLE-based method for estimating the coefficients.  ML-APA is a block update algorithm that updates its estimate every $P$ samples. In Section \ref{sec:iml}, we derive an alternative MLE-based method, incremental maximum likelihood APA (IML-APA) that incrementally updates its estimate at every time step. We will show that although IML-APA is harder to analyze than ML-APA, it is closely related and empirically performs better in some scenarios of interest. Section \ref{sec:simulations} presents our simulation results that show the effectiveness of ML-APA and IML-APA methods in AEC. The proof of our main theoretical result is presented in Section \ref{sec:proof}. Section \ref{sec:conclude} concludes the paper. 

\section{Statistical  behavior of APA}\label{sec:apa}
APA is an algorithm that was proposed to address the slow convergence problem of LMS and NLMS algorithms. Unlike LMS and NLMS, APA employs $P$ latest observation vectors, for some $P>1$.  Let $X_t = [\xv_t,\ldots,\xv_{t-P+1}]$, and $\yv_t=[y_t,\ldots,y_{t-P+1}]^T$. The standard APA algorithm is designed as follows: Given $(X_t,\yv_t)$ and $\wv_t$, the coefficients are updated such that the new vector is closest vector in $\mathds{R}^{L}$, which satisfies the last $P$ measurements exactly, i.e., $X_t^T\wv=\yv_t$. That is, it sets $\wv_{t+1}=\arg\min_{\wv:\;X_t^T\wv=\yv_t.} \|\wv-\wv_t\|^2$, or  
\[
\wv_{t+1}=\wv_{t} + X_t(X_t^TX_{t})^{-1}(\yv_t-X_t^T\wv_t).
\] 
APA in its standard form shows  instability issues that are caused by the inversion involved in $(X_t^TX_{t})^{-1}$. To address this issue, regularized APA is defined as 
\[
\wv_{t+1}=\wv_{t} + \mu X_t(X_t^TX_{t}+\delta I_P)^{-1}(\yv_t-X_t^T\wv_t),
\]
where $\delta$ and $\mu$ denote the regularization parameter and step size, respectively. 
Regularized APA can be motivated analytically as follows.
 First, note that, because the measurements are noisy, even the true coefficient vector $\wv^*$ does not  satisfy  $X_t^T\wv=\yv_t$. The noise variance determines the typical distance to be expected between $X_t^T\wv$ and $\yv_t$. Taking noise into account, we could tweak the APA optimization as follows:
\begin{align}
\min \;\;\;& \|\wv-\wv_t\|^2\nonumber\\
{\rm s.t.}\;\;\;& \|X_t^T\wv-\yv_t\|^2\leq P\sigma_z^2.
\end{align}
Solving  the Lagrangian form of the above optimization $\min_{\wv} [\|\wv-\wv_t\|^2+\lambda  \|X_t^T\wv-\yv_t\|^2]$, it follows that
\begin{align}
\wv_{t+1}= (I+\lambda X_tX_t^T)^{-1}(\wv_t+\lambda X_t\yv_t).\label{eq:form1}
\end{align}
Employing Woodbury matrix identity, we can simplify this update rule:
\begin{align}
\wv_{t+1}
&= (I+\lambda X_tX_t^T)^{-1}(\wv_t+\lambda X_t\yv_t)\nonumber\\
&=(I-\lambda X_t(I_P+\lambda X_t^TX_t)^{-1}X_t^T)(\wv_t+\lambda X_t\yv_t)\nonumber\\
&=\wv_t +  X_t(\delta I_P+ X_t^TX_t)^{-1}( \yv_t- X_t^T\wv_t),\label{eq:form2}
\end{align}
where $\delta=1/\lambda$.

To understand the convergence behavior of APA, we employ the update rule in \eqref{eq:form1}. Recall that under our model, $\yv_t=X_t^T\wv^*+\zv_t$,
where  $\zv_t=[z_t,\ldots,z_{t-P+1}]$. Subtracting $\wv^*$ from both side of \eqref{eq:form1}, it follows that 
\begin{align}
\wv_{t+1}-\wv^*=&\left(I-X_t(\delta I_P+ X_t^TX_t)^{-1}X_t^T \right)(\wv_{t}-\wv^*)\nonumber\\
&+ X_t(\delta I_P+ X_t^TX_t)^{-1}\zv_t.
\end{align}
Assume that $\wv_0={\bf 0}_{L}$,  and define
\[
P_t=I_L-X_t(\delta I_P+ X_t^TX_t)^{-1}X_t^T. 
\]
it follows that $\wv_{t+1}=\wv^*-\prod_{t'=1}^t P_{t'} \wv^*+\sum_{t'=1}^t\prod_{t''=t'+1}^t P_{t''} X_{t'}(\delta I_P+ X_{t'}^TX_{t'})^{-1}Z_{t'}$.
This shows that at time $t+1$ conditioned on the input signal, $\wv_{t+1}$ has a Gaussian distribution with mean
\[
\E[\wv_{t+1}]=\wv^*-\prod_{t'=1}^t P_{t'}  \wv^*. 
\]
Note that  $P_t$ has $L-P$ eigenvalues that are equal to one. The corresponding  eigenvectors represent the directions that are orthogonal to the space spanned by the columns of $X_t$. For directions that are not orthogonal to the space spanned by the columns of $X_t$, the corresponding eigenvalues $\nu_i$ are strictly smaller than $1$, and in fact can be expressed as $\nu_i = \delta/\left(\delta + S_{ii}^2\right)$ where $\{S_{ii}\}$  are the non-zero singular values of $X_t$.  This implies that the estimation bias along the directions that are observed in the input data vanishes quickly, as long as the regularization parameter $\delta$ is not larger than the typical singular values of $X_t$.


\section{ML-APA: Online MLE  for adaptive filtering}\label{sec:ML-alg}

Our analysis of the previous section suggests that $\wv_t$, the APA's estimate of $\wv^*$  at iteration $t$, is a Gaussian random vector with a mean quickly converging  to $ \wv^*$. In this section, first inspired by the analysis of the previous section, we derive a new AEC algorithm that employs MLE. For i.i.d.~Gaussian inputs, we fully characterize the expected temporal performance of the proposed method and show that it converges to the optimal coefficient at a rate that is asymptotically optimal. 

Assume that at time $t$ we have access to i) $\wv_t\sim \Nc(\wv^*,M_t)$, for some covariance matrix $M_t$, and ii) $(X_t,\yv_t)$, where $X_t\in\mathds{R}^{L\times P}$ and $\yv_t\in\mathds{R}^P$, and $\yv_t=X_t^T\wv^*+\zv_t$. 
Further, assume that $\wv_t$ is independent of $(X_t,\yv_t)$; it however may be a function of ($X_{t-P},\yv_{t-P}$). To support this assumption, we will derive an ML based estimate that performs block updates, i.e. that moves forward in steps of $P$ samples.
Given $\wv_t$ and $(X_t,\yv_t)$, one can ask the following question: what is the ML estimate of $\wv^*$? Given the distributions of $\wv_t$  and $(X_t,\yv_t)$, the ML estimate $\wv^*$ is the solution of the following optimization:
\[
\wv^{\rm (ML)}_{t+P}=\argmin_{\wv\in\mathds{R}^L}[ (\wv_t-\wv)^TM_t^{-1}(\wv_t-\wv) +{1\over \sigma_z^2}\|\yv_t-X_t^T\wv\|^2].
\]
Therefore, 
\begin{align}\label{eq:def-ml}
\wv^{\rm (ML)}_{t+P}=(M_t^{-1}+{1\over \sigma_z^2}X_tX_t^T)^{-1}({1\over \sigma_z^2}X_t\yv_t+M_t^{-1}\wv_t).
\end{align}
Note that 
\begin{align}
\wv^{\rm (ML)}_{t+P}
&=(\sigma_z^2M_t^{-1}+X_tX_t^T)^{-1}\Big(X_t(\yv_t-X_t^T\wv_t+X_t^T\wv_t)\nonumber\\
&\;\;\;\;\;+\sigma_z^2M_t^{-1}\wv_t\Big)\nonumber\\
&=(\sigma_z^2M_t^{-1}+X_tX_t^T)^{-1}\Big(X_t(\yv_t-X_t^T\wv_t)\nonumber\\
&\;\;\;\;\;+(X_tX_t^T+\sigma_z^2M_t^{-1})\wv_t\Big)\nonumber\\
&=\wv_t+(\sigma_z^2M_t^{-1}+X_tX_t^T)^{-1}X_t(\yv_t-X_t^T\wv_t).
\end{align}
To simplify the update rule, assume that the distribution of $\wv_t$ is homogenous, i.e.,  $M_t = m_t I_L$, for some scalar $m_t$. Using the identity $(I+AB)^{-1}A=A(I+BA)^{-1}$,  $(\sigma_z^2 M_t^{-1}+X_tX_t^T)^{-1}X_t = X_t \left( c^{-1}I + X_t^T X_t\right)^{-1}$, 
where, $c_t$,  is defined as $c_t= {m_t\over \sigma_z^2}.$ 
Then, $\wv^{\rm (ML)}_{t+P}$ can be written as $\wv^{\rm (ML)}_{t+P}=\wv_t+ X_t({1\over c_t}I_P+  X_t^TX_t)^{-1}(\yv_t-X_t^T\wv_t).$ 
This expression provides the MLE of $w^*$ given the assumption $\wv_t\sim \Nc(\wv^*,m_t I)$ with known parameter $m_t$.
Moreoever, if we know that the covariance of $\wv_t$ is of the form $m_t I$, but $m_t$ is unknown, we can derive the MLE of $m_t$ as the solution to the optimization
\[
m^{({\rm ML})}_t=\argmax_{m\in\mathds{R}^+} \Big( -{L\over 2}\log m-{1\over 2m}\|\wv_t-\wv^*\|^2\Big),
\]
which leads to 
\[
m^{({\rm ML})}_t={1\over L}\|\wv_t-\wv^*\|^2,
\] 
often referred to as the misalignment in echo cancellation literature.
Putting the pieces together, we derive the update rule which we refer to as ML-APA update rule: 
\begin{align}
\wv_{t+P}=\wv_t+ X_t({1\over c_t}I_P+  X_t^TX_t)^{-1}(\yv_t-X_t^T\wv_t),\label{eq:def-ga-obml}
\end{align}
with the \emph{misalignment to noise ratio (MNR)} defined as 
\begin{align}
c_t= {\|\wv_t-\wv^*\|^2 \over \sigma_z^2L}.\label{eq:def-ct}
\end{align}
To compute the MNR at time $t$, one needs to have access to the misalignment at time $t$, i.e., $\|\wv_t-\wv^*\|^2$, which is obviously not available in practice. That is why we will refer to this version of the algorithm  as ``oracle-aided' ML-APA, or OA-ML-APA. However, as explained in Section \ref{sec:simulations}, MNR can be estimated reasonably well in practice. In particular,  the  ``delay and extrapolation'' approach \cite{hansler2005acoustic} enables us to estimate  the misalignment  with only marginal loss of performance compared with the oracle-aided version in some example scenarios with dynamic channel changes. Pseudo-code for the OA-ML-APA algorithm is provided in Algorithm \ref{alg:ga-obml} .

\begin{algorithm}
\caption{OA-ML-APA algorithm}\label{alg:ga-obml}
\begin{algorithmic}
\Require $(\xv_1,\ldots,\xv_T)$, with $\xv_t\in\mathds{R}^L$, $(y_1,\ldots,y_T)$, $\wv^*\in\mathds{R}^L$, $P$, $L$ and $\sigma_z$
\Ensure $\wv$
\State{Initialize $\wv\gets {\bf 0}_L$}
\For{$t \gets 1$ to $T$,  step = $P$}
    \State{ $X\gets [\xv_t,\ldots,\xv_{t-P+1}]$}
    \State{ $\yv \gets [y_t,\ldots,y_{t-P+1}]^T$}
  \State{ $c\gets \|\wv-\wv^*\|^2 /(L\sigma_z^2)$ }
  \State{ $\ev\gets \yv-X^T\wv$}
  \State{ $\wv\gets \wv+ X({1\over c}I_P+  X^TX)^{-1}\ev$}
\EndFor
\end{algorithmic}
\end{algorithm}

Some intuitive insight into the ML-APA algorithm can be obtained by considering different ranges of the MNR ($c_t$).
 During initial steps, $\|\wv_t-\wv^*\|^2$ is large and therefore $c_t$ is large as well. In this case, ML-APA behaves more like standard APA with aggressive step size $\mu = 1$ and small regularization $\delta$. As time proceeds and the misalignment error becomes small, the MNR $c_t$ becomes small as well. In that case, ML-APA  becomes more conservative and its update rule becomes similar to block LMS with a very small step size $\mu = c_t$. That is, we have approximately $\wv_{t+P} = \wv_t + c_t X_t \left( \yv_t - X_t^T \wv_t \right) $. Therefore, in summary, by setting the MNR as \eqref{eq:def-ct}, ML-APA is able to perform a soft transition between APA and block-LMS, depending on the current ratio of misalignment and noise. 

To derive the ML-APA update rule we made several assumptions, such as unbiasedness of our estimates, which, at best, are approximately true. However, our next theorem, which is our main theoretical result, does not depend on these assumptions, fully characterizes the expected misalignment performance of OA-ML-APA, and shows that it indeed is able to asymptotically recover the unknown coefficients $\wv^*$. 
It is convenient to define the normalized misalignment $a_t = \| \wv_t-\wv^* \|^2 \sigma_x^2/\sigma_z^2$.
\begin{theorem}\label{thm:main}
Assume that $\yv_t=X_t^T\wv^*+\zv_t$, where $X_t\in \mathds{R}^{L\times P}$ and $\zv_t\in\mathds{R}^{P}$ are independent and the entries of $X_t$ and $\zv_t$ are  i.i.d.~$\Nc(0,\sigma_x^2)$ and  $\Nc(0,\sigma_z^2)$, respectively. Let  $\wv_{t+P}=\wv_t+ X_t({1\over c_t}I_P+  X_t^TX_t)^{-1}(\yv_t-X_t^T\wv_t)$, where $\wv_t\in\mathds{R}^{L}$ is a fixed vector and parameter $c_t$ is defined as \eqref{eq:def-ct}. Let $L$ be large enough that $\sqrt{L} - \sqrt{\log L} > \sqrt{P}$. Then, 
\begin{align}
\E[a_{t+P}]\nonumber
&\leq  (1-  (1-\gamma_L)  \frac{P}{L} \frac{ a_t }{1 +  a_t }) a_t
\end{align}
where $\gamma_L=O(  \sqrt{\frac{\log L}{L}} )$ is an absolute constant depending only on 
$P$ and $L$. 
\end{theorem}
The proof of Theorem \ref{thm:main} is provided in Section \ref{sec:proof}, where we show that
\[
\gamma_L = 1 - (1 - \sqrt{\frac{P}{L}} - \sqrt{\frac{\log L}{L}})^2(1-\frac{2}{\sqrt{L}} ).
\]

Theorem \ref{thm:main} shows that after every update the expected misalignment at time $t+P$ (namely $\E[a_{t+P}]$) is smaller than the actual  error at time $t$ (namely $a_t$) by a factor that is smaller than one but depends on $a_t$. To further understand the implications of Theorem \ref{thm:main}, define function $f:\mathds{R}\to\mathds{R}$ as
\[
f(a)=\Big(1-\eta {a\over 1+a}\Big)a,
\]
where $\eta\in(0,1)$ is defined as
\[
\eta=\left(1-\gamma_L\right) {P\over L}.
\] 
Using these definitions, Theorem \ref{thm:main} states that 
\begin{align}
\E[a_{t+P}]\leq f(a_t),\label{eq:bound-at-new}
\end{align}
 for  $t=0,P,2P,\ldots$.
Since $f$ is a concave function, 
taking the expected value of both side of  \eqref{eq:bound-at-new} and using  Jensen's inequality, it follows that
\begin{align}
\E[a_{t+P}]&\leq f\left( \E[a_t]\right) \nonumber\\
&= (1-\left(1-\gamma_L\right) ({P\over L}){\E[a_t]\over 1+\E[a_t]})\E[a_t].
\end{align}
Thus if we repeatedly apply ML-APA, the sequence defined by  $a_{t+P} = f(a_t)$ is an upper bound on the expected value of the normalized misalignment after 
$t$ samples. 
This expected error reduces monotonically, and as we will see, asymptotically achieves the performance achieved by   optimal offline algorithms.

For further insight into the effectiveness of the proposed  ML-APA method, we next compare its performance, characterized in Theorem \ref{thm:main}, with a lower bound on the performance achieved  by a general class of offline algorithms that estimate $\wv^*$ at time $s$ using the entire history of observations.
Let $\yv_s=X_s^T\wv^*+\zv_s$, where $X_s\in \mathds{R}^{L\times s}$ and $\zv_s\in\mathds{R}^{s}$. We define $\wv^{(\delta)}_{s}$, 
a generalized regularized least squares estimator, defined as 
\[
\wv^{(\delta)}_{s}=\wv_0+\Delta \wv,
\] 
where $\wv_0$  denotes  a rough estimate of the desired parameters ($\wv^*$) available from a different set of data.  Moreover, 
\[
\Delta \wv=\arg\min_{\wv} [\|X_s^T\wv_0+X_s^T\wv-\yv_s\|^2+\delta \|\wv\|^2].
\]  
Solving for $\Delta \wv$, it is straightforward to see that
$$
\wv^{(\delta)}_{s}=\wv_0+ X_s(\delta I_s+  X_s^TX_s)^{-1}(\yv_s-X_s^T\wv_0).
$$
Note that setting $\delta = 0$ and assuming that $s\geq L$, $\wv^{(0)}_{s}=X_s(\delta I_s+  X_s^TX_s)^{-1}\yv_s$  denotes the  ordinary least squares estimate of $\wv^*$.  In that special case, $\wv^{(0)}_{s}$ is  an unbiased estimate of $\wv^*$, which is the minimum variance unbiased estimator under our assumption that  $Z_s$ is Gaussian. With $\delta > 0$, the estimator is no longer unbiased, but if $\delta$ is chosen correctly, the mean squared error can be somewhat lower than that of the unbiased estimate. Theorem \ref{thm:offline} below provides a lower bound on the mean squared error achieved by $\wv^{(\delta)}_{s}$ and shows that if the initial error $a_0=\|\wv_0-\wv^*\|^2\sigma_x^2/\sigma_z^2$ is {\em known}, the best regularization factor parameter  is 
\[
\delta = c_0^{-1} = {L\sigma_x^2\over a_0}.
\] 
(Note also that $\wv^{(c_0^{-1})}$ happens to be the ML estimate of $\wv^*$, under a Bayesian model with $\wv_0 \sim{\cal{N}}({\bf 0}_L,\frac{a_0}{L} I_L )$ .)

\begin{theorem}\label{thm:offline}
Assume that $\yv_s=X_s^T\wv^*+Z_s$, where $X_s\in \mathds{R}^{L\times s}$ and $Z_s\in\mathds{R}^{s}$ are independent and the entries of $X_s$ and $Z_s$ are  i.i.d.~$\Nc(0,\sigma_x^2)$ and  $\Nc(0,\sigma_z^2)$, respectively. Let  $\wv^{(\delta)}_{s}=\wv_0+ X_s(\delta I_s+  X_s^TX_s)^{-1}(\yv_s-X_s^T\wv_0)$, where $\wv_0\in\mathds{R}^{L}$ is a fixed vector and $\delta \geq 0$ .  Let $a^{(\delta)}_s=\|\wv^{(\delta)}_{s}-\wv^*\|^2\sigma_x^2/\sigma_z^2$  . Then, if $s \leq L$,
\begin{align}
\E[a^{(\delta)}_{s}]  \nonumber
&\geq \E[a^{(c_0^{-1})}_{s}]  \geq  \left(1-    \frac{s}{L} \frac{  a_0 }{1+  a_0 }\right) a_0,
\end{align}
and if $s \geq L$,
\begin{align}
\E[a^{(\delta)}_{s}]  \nonumber
&\geq \E[a^{(c_0^{-1})}_{s}]  
\geq \frac{L a_0}{s a_0 + L}.
\end{align}
\end{theorem}

Theorem \ref{thm:offline} provides a lower bound on any regularized least squares method for offline estimation of $\wv^*$ using $s$ samples. 
(In the case of $\delta = 0$, $s<L$, the matrix inverse in the definition of $\wv^{(\delta)_s}$ should be interpreted as a pseudo-inverse.)

 The bound has two phases - an initial fast convergence from $a_0$ to $a_0/(1+a_0) \approx 1$, after $s = L$ samples, followed by a gradual improvement with $a_s  \approx L/s$, as $s\to\infty$.
Below, we examine the implications of Theorems \ref{thm:main} and \ref{thm:offline} in these two regimes
 and show that despite being an on-line algorithm with fixed memory $P$, ML-APA performs nearly as well as offline methods, for memoryless inputs.

Next we  compare the upper bound on ML-APA performance (the sequence $a_t$ generated  as $a_{t+P} = f(a_t)$) with   the lower bound  $a_s$ on performance of offline regularized least squares.



\subsection{Initial steps}\label{sec:opt-ga-obml-init}
Assume that we initialize ML-APA algorithm at $\wv_0={\bf 0}_L$, which corresponds to  $a_0=\|\wv^{*}\|^2\sigma_x^2/\sigma_z^2$. 
If the filter $\wv^*$ is such that initial residual echo dominates the noise, we have $a_0 \gg 1$, which implies that 
${a_t\over 1+a_t}\approx 1$ and therefore, during initial steps, we expect ML-APA to show an almost linear convergence with $a_{t+P}\leq (1-{P\over L})a_t$ or $a_{t}\leq (1-{P\over L})^{t/P} a_0 \approx e^{- (P/L) (t/P)} a_0 = e^{-t/L} a_0$, where we have assumed $P \ll L$.
ML-APA thus takes approximately $L \log a_0$ samples to achieve $a_t \approx 1$, just a logarithmic factor longer than the $L$ samples required by the offline estimation.

Note that in this analysis, for memoryless input,  the convergence speed (in terms of the number of samples) does not depend on $P$. The motivation for  studying memory order $P>1$ is that increasing $P$ improves convergence rate for correlated sources, such as those observed in audio echo cancellation  applications. Later we show empirically that for audio input signals the convergence rate of ML-APA improves with $P$.

\subsection{Asymptotic optimality of ML-APA}\label{sec:opt-ga-obml}
Next we compare the asymptotic  convergence behavior  of ML-APA with our lower bound on offline regularized least squares methods. 

From Theorem \ref{thm:offline}, for $s \geq L$, the normalized expected error of offline algorithms is at least $a_s  = L a_0/( s a_0 + L)$, or $a_s \approx L/s$.

For ML-APA,  recall that the upper bound sequence $a_{t+P}= a_t(1-(1-\gamma_L)({P\over L}){a_t\over 1+a_t})$ is defined for $t = kP$. Changing notation to $\tilde{a}_k := a_{kP}$, we show in Lemma \ref{lemma:app-b}  in Appendix \ref{app:b}  that for large enough $k$,  $\tilde{a}_k \approx (1-\gamma_L)^{-1} L/(Pk)$.  Thus
$a_t \approx (1-\gamma_L)^{-1} L/t$,
demonstrating that ML-APA achieves asymptotically the same performance as OLS, up to a factor $(1-\gamma_L)^{-1}$ that approaches 1 for large $L$.

\begin{figure}[h]
\centering
        \includegraphics[width=0.45\textwidth]{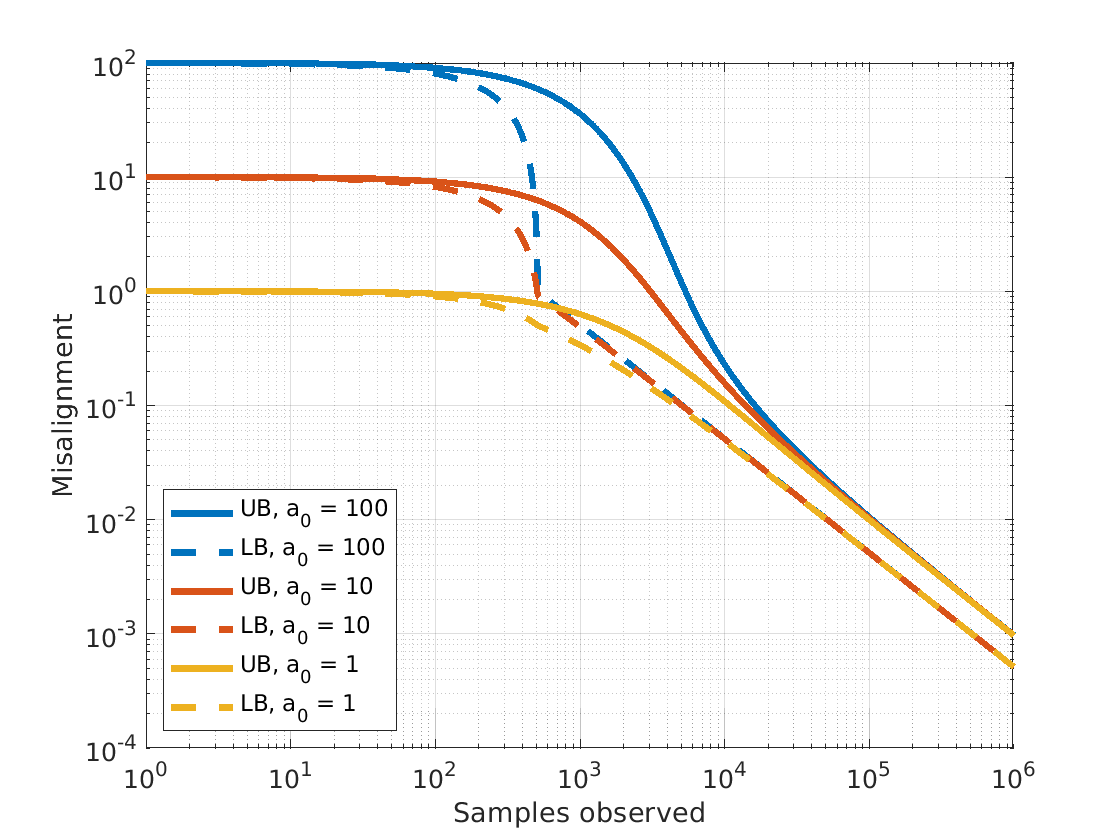}
         \caption{Comparing bounds on misalignment of OA- ML-APA to bounds on misalignment of offline methods, for three different initial values. (Here, $L=512$ and $P=8$.)}
         \label{fig:at}
\end{figure}

Fig.~\ref{fig:at} shows the temporal behavior of upper and lower bounds for $a_t$ derived for ML-APA and offline methods, respectively, for various initial error levels $a_0$, and for $L=512$ and $P=8$. The performances of the two methods  coincide for $t\ll L$ and differ only by the factor $1-\gamma_L$, for $t \gg L$. At the transition between the two regimes ($t\approx L$),  a gap between the bounds on on-line and offline methods is apparent. It is striking that the initial level of error $a_0$ does not affect the long-term behavior - scenarios with larger initial error are able to "catch up" to scenarios with lower initial error. In Section~\ref{sec:simulations}, the bounds are compared with empirically evaluated performance of OA-ML-APA and OLS.


\section{Incremental ML APA}\label{sec:iml}

In Section \ref{sec:ML-alg}, we proposed and studied  ML-APA, an adaptive MLE-based echo cancellation algorithm that updates its estimate of the echo filter coefficients every $P$ samples. Most existing echo cancellation algorithms on the other hand, update their estimates  every  time sample. In this section,  inspired by ML-APA, we propose an alternative solution, which updates its online ML estimate of the parameters at every time step.  We refer to the new solution as the incremental maximum likelihood (IML) APA (IML-APA) algorithm. As we will show in Section \ref{sec:simulations}, in practice, the performance of IML-APA in general  is as good as that of ML-APA.

Recall that  $X_t = [\xv_t,\ldots,\xv_{t-P+1}]$, and $\yv_t=[y_t,\ldots,y_{t-P+1}]^T$. Let 
\[
U_{t} = [\xv_t,\ldots,\xv_{t-P+2}],
\]
and $\vv_t = [y_{t},\ldots,y_{t-P+2}]$. Using these definitions, 
\[
X_t=[\xv_t,U_{t-1}], \;\;\; \yv_t=[y_t,\vv_{t-1}^T]^T.
\]
Assume that at time $t$, we are given $\wv_{t-P}$ which is an unbiased  estimate of $\wv^*$, such that $\wv_{t-P} \sim\Nc(\wv^*,m_{t-P} I_L)$. Then,  we compute two possible updates: i) $\wv_1$: ML estimate of $\wv^*$ given  $\wv_{t-P}$ and $P-1$ new observations $(U_{t-1},\vv_{t-1})$ , and ii)  $\wv_2$: ML estimate of $\wv^*$ given  $\wv_{t-P}$ and $P$ new observations  $(X_t,\yv_t)$ . From (\ref{eq:def-ml}), we have 
\begin{align} \label{eq:w1}
\wv_1=(c^{-1}I_L+U_{t-1}U_{t-1}^T)^{-1}(\wv_{t-P}+U_{t-1}\vv_{t-1}),
\end{align}
and 
\begin{align}\label{eq:w2}
\wv_2=(c^{-1}I_L+X_tX_t^T)^{-1}(\wv_{t-P}+X_t\yv_t),
\end{align}
where $c= { m_{t-P}\over \sigma_z^2 }$. 
But, $X_tX_t^T=U_{t-1}U_{t-1}^T+\xv_t\xv_t^T$ and $X_t\yv_t=U_{t-1}\vv_{t-1}+y_t \xv_t$. Let $B_t=c^{-1}I_L+U_{t-1}U_{t-1}^T$. Using the Woodbury matrix identity, 
\[
(c^{-1}I_L+X_tX_t^T)^{-1}=B_t^{-1}-{B_t^{-1}\xv_t\xv_t^TB_t^{-1}\over 1+\xv_t^TB_t^{-1}\xv_t}. 
\]
Therefore, $\wv_1$ and $\wv_2$ can be connected as
\begin{align}
\wv_2&=\left(B_t^{-1}-{B_t^{-1}\xv_t\xv_t^TB_t^{-1}\over 1+\xv_t^TB_t^{-1}\xv_t}\right)(\wv_t+U_{t-1}V_{t-1}+y_t\xv_t)\nonumber\\
&=\left(I_L-{B_t^{-1}\xv_t\xv_t^T\over 1+\xv_t^TB_t^{-1}\xv_t}\right)(\wv_1+y_tB_t^{-1}\xv_t)\nonumber\\
&=\wv_1+{B_t^{-1}\xv_t(y_t-\xv_t^T\wv_1)\over 1+\xv_t^TB_t^{-1}\xv_t}.\label{eq:connection-w2-w1}
\end{align}
Thus if we had access to the ML estimate of $\wv^*$ based on the previous $P-1$ samples ($\wv_1$), given the new observation point $(\xv_t,y_t)$, we can use \eqref{eq:connection-w2-w1} to obtain the ML estimate based on $P$ samples ($\wv_2$) incrementally, without reference to $\wv_{t-P}$.  This leads to our proposed IML-APA update rule. At time $t$, we assume that our current estimate of the filter coefficients $\wv_t$ is  the ML estimate of $\wv^*$ based on the previous $P-1$ samples and $\wv_{t-P}$. Then, given new data $(\xv_t,y_t)$ we design $\wv_{t+1}$ to be the ML estimate based on the past $P$ samples and $\wv_{t-P}$. That is, 
\begin{align}
\wv_{t+1}&=\wv_t+{B_t^{-1}\xv_t(y_t-\xv_t^T\wv_t)\over 1+\xv_t^TB_t^{-1}\xv_t},\label{eq:IML-long}.
\end{align}
Intuitively, this update rule extracts relevant information from the  latest sample, assuming that the previous $P-1$ samples have already been used effectively, in a maximum likelihood framework.  Defining $\tilde{\xv}_t = c_t^{-1} B_t^{-1}\xv_t $, we can also write
\begin{align}
\wv_{t+1}&=\wv_t+{ \tilde{\xv}_t(y_t-\xv_t^T\wv_t) \over c_t^{-1}+\xv_t^T\tilde{\xv}_t},\label{eq:IML-NLMS}.
\end{align}
In this form, the update is seen to be similar to a VR-NLMS update, but with $\tilde{\xv}_t$ replacing $\xv_t$ in some places.
Computation of $\tilde{\xv}_t$ only requires a matrix inverse of size $P-1$, since again by the 
Woodbury matrix identity, 
\begin{align}
B_t^{-1}=cI_L-cU_{t-1}(c^{-1}I_{P-1}+U_{t-1}^TU_{t-1})^{-1}U_{t-1}^T. \nonumber
\end{align}
IML-APA's update rule can also  be written in an alternate form closer to ML-APA's.
Note that
$$
(c^{-1}I_L + X_t X_t^T)\frac{ B_t^{-1} \xv_t}{1 + \xv_t^T B_t^{-1} \xv_t} = \frac{(B_t +\xv_t \xv_t^T)B_t^{-1} \xv_t}{1 + \xv_t^T B_t^{-1} \xv_t} = \xv_t
$$
so that (\ref{eq:IML-long}) becomes
\begin{align}
\wv_{t+1}&=\wv_t+ \left( c^{-1}I_L + X_t X_t^T \right)^{-1} \xv_t (y_t-\xv_t^T\wv_t) \nonumber \\
& = \wv_t+ \left( c^{-1}I_L + X_t X_t^T \right)^{-1} X_t 
\left[
\begin{array}{c}
  \epsilon_t  \\
{\bf 0}_{P-1} 
\end{array}
\right] \nonumber \\
& = \wv_t+ X_t \left( c^{-1}I_P + X_t^T X_t \right)^{-1} 
\left[
\begin{array}{c}
  \epsilon_t  \\
{\bf 0}_{P-1} 
\end{array}
\right]  \label{eq:IML-APA}
\end{align}
where $\epsilon_t=y_t-\wv_t^T\xv_t$. In this form, the IML-APA update rule is very similar to that of ML-APA, except that in IML-APA, the filter coefficient estimates are updated after every sample, instead of every $P$ samples, and all but the first term in the error vector are zeroed out. 
Pseudocode fo oracle-aided OA- IML-APA in this form is provided in Algorithm \ref{alg:ga-iml}.

\begin{algorithm}
\caption{OA-IML-APA algorithm}\label{alg:ga-iml}
\begin{algorithmic}
\Require $(\xv_1,\ldots,\xv_T)$, with $\xv_t\in\mathds{R}^L$, $(y_1,\ldots,y_T)$, $\wv^*\in\mathds{R}^L$, $P$, $L$ and $\sigma_z$
\Ensure $\wv$
\State{Initialize $\wv\gets {\bf 0}_L$}
\For{$t \gets 1$ to $T$}
    \State{ $X\gets [\xv_t,\ldots,\xv_{t-P+1}]$}
    \State{ $\yv \gets [y_t,\ldots,y_{t-P+1}]^T$}
  \State{ $c\gets \|\wv-\wv^*\|^2 /(L\sigma_z^2)$ }
  \State{ $e\gets y_t-\xv_t^T\wv$}
  \State
  \State{ $\ev\gets 
\left[
\begin{array}{c}
e  \\
  {\bf 0}_{P-1}
  \end{array}
\right]$}
\State
  \State{ $\wv\gets \wv+ X({1\over c}I_P+  X^TX)^{-1}\ev$}
\EndFor
\end{algorithmic}
\end{algorithm}


\section{Simulation results}\label{sec:simulations}

Our goal in this section is to compare the ML-APA and IML-APA methods with each other, with existing adaptive filtering methods, and with the analytical results of this paper in a simulated acoustic echo cancellation setting.  The first simulation uses a white noise input signal to be able to compare with analytical results, while the subsequent simulations are based on correlated auto-regressive processes or on recorded audio signals as input.  We also introduce and evaluate practical versions of IML-APA and ML-APA,  in which the misalignment $\|\wv_t-\wv^*\|^2$ is estimated by a delay-and-extrapolate technique rather than provided by an oracle. 

For the simulations with recorded audio input signals, we have used the publicly available  audio dataset Librispeech, described  in  \cite{panayotov2015librispeech}. For every simulation result, we performed $10$ trials using randomly selected audio segments from this dataset, and reported  the average performance. The signals are sampled at $f_s= 16$ kHz. In most simulation results the echo channel response is of length $L=512$ corresponding to $32$ (ms). In the misalignment plots, we plot the normalized misalignment in dB, $10 \log_{10}(a_t)$. 

In deriving our theoretical results, we assumed that the input vectors $\xv_t$ are independent from each other, and also that the elements within a given vector are uncorrelated.  For AEC, (and related applications), neither of these assumptions typically  hold, as consecutive vectors have considerable overlap, and also consecutive time samples are strongly correlated. (Recall that $\xv_{t+1}=[x_{t+1},x_t,\ldots,x_{t-L+2}]$ and $\xv_{t}=[x_{t},\ldots,x_{t-L+1}]$.) However, in our simulation results, we show that the ML-APA and IML-APA methods   achieves state-of-the-art performance, even in AEC scenarios where the input vectors are not independent and have overlap.

\subsection{Performance under i.i.d.~input}\label{sim:iid}

To illustrate directly the consequences of  Theorems~\ref{thm:main} and \ref{thm:offline}, we first simulated an echo cancellation scenario in which the far-end signal is an iid Gaussian sequence. 

\begin{figure}[h]
\centering
        \includegraphics[width=0.45\textwidth]{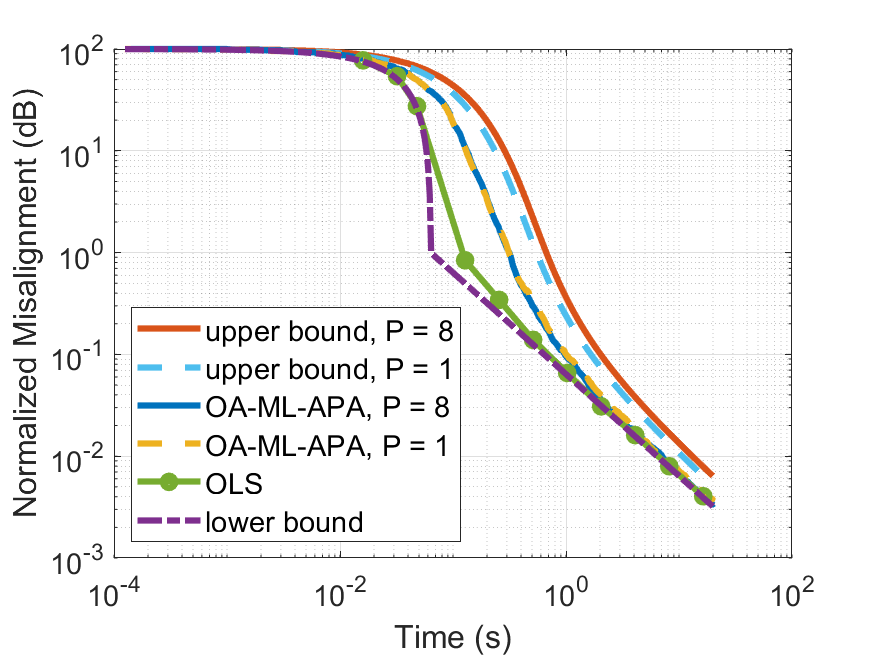}
         \caption{Comparing performace of OA-ML-APA and OLS with theortical bounds, for $L=512$ and two values of $P$}
         \label{fig:oaml_vs_theory}
\end{figure}

We simulated a linear time invariant system \eqref{eq:lti_system} where the input sequences $x_t$ and $z_t$ are i.i.d. Gaussian with variances 1 and 0.01, respectively,  and where the fixed channel vector $\wv^*$ of length $L = 512$ is generated with i.i.d. Gaussian coefficients and normalized to $\| \wv^* \| = 1$, and our estimate of the filter coefficients is initialized to  $\wv_0 = {\bf 0}_L$. With these parameters, the initial normalized misalignment is $a_0 = 100$.
From this initialization, we ran the OA-ML-APA algorithm for 160,000 samples, once with projection order $P = 1$ and once with $P = 8$, recording the normalized misalignment $a_s$ at each time sample.  We also computed the offline least squares solution with fixed optimized regularization $c_0^{-1}$, performing the calculation only at logarithmically-space time intervals to reduce complexity. The evolution of $a_s$ over time in these three scenarios is plotted, together with the upper and lower bounds from Theorems~\ref{thm:main} and \ref{thm:offline}, in Figure~\ref{fig:oaml_vs_theory},  assuming a sampling rate of 8 kHz. As predicted by the theory, the performance of the three filtering methods lie between the bounds. Interestingly, in this i.i.d. example, the performance of OA-ML-APA does not depend materially on $P$, and OA-ML-APA coincides with OLS and with the OLS lower bound after about two seconds (16,000 samples).

\subsection{Impact of input correlation}\label{sim:corr_input}

\begin{figure}[h]
\centering
        \includegraphics[width=0.45\textwidth]{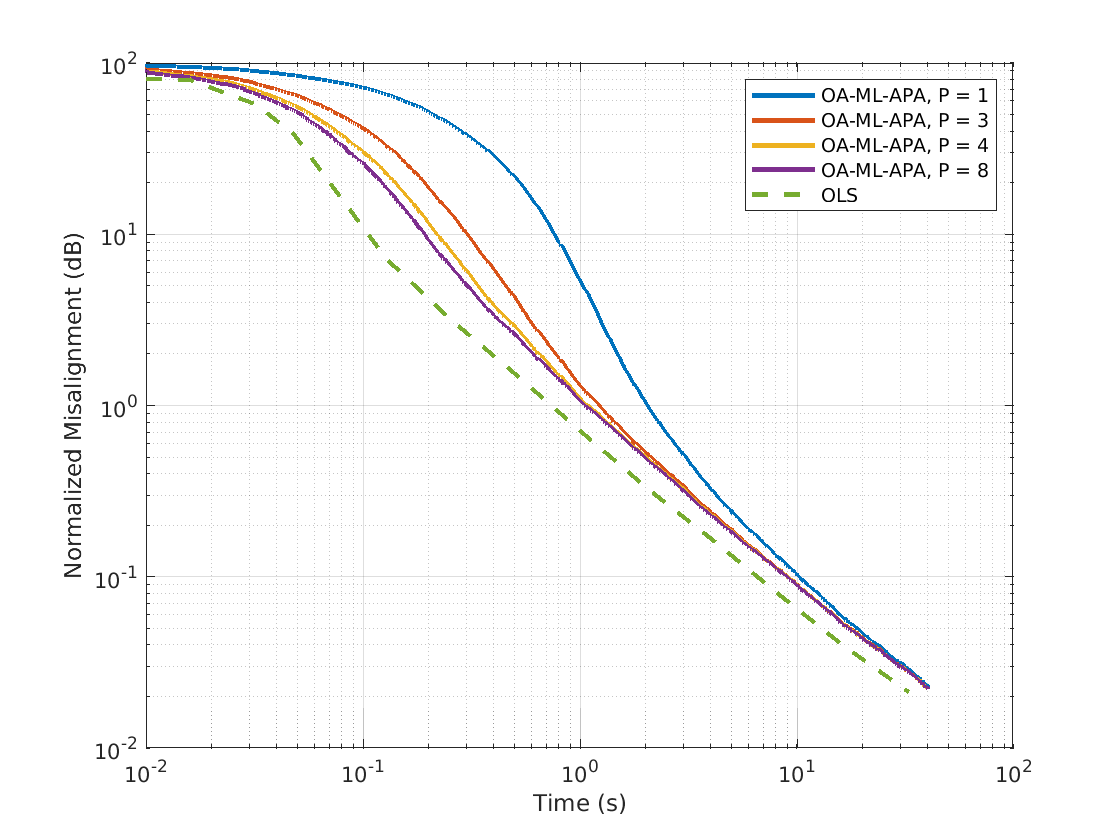}
         \caption{Comparing performace of OA-ML-APA and OLS when the input is an autoregressive process, for $L=512$ and for $P=1,2,4,8$}
         \label{fig:oaml_vs_ols_with_ar}
\end{figure}

In audio applications, input signals are typically strongly correlated, in which case our theoretical bounds no longer apply. To study the effects of a correlated input in the simplest setting, we replaced the i.i.d. input $z(t)$  of Section~\ref{sim:iid} with an autoregressive process $\tilde{z}(t) = \rho \tilde{z}(t-1) + \sqrt{1-\rho^2}  z(t)$. For each curve, we performed 10 independent trials and averaged the misalignment results.  The theoretical bounds don't apply in this case. Figure~\ref{fig:oaml_vs_ols_with_ar} shows that both OLS and OA-ML-APA converge more slowly in the presence of correlated input, but that the long-term performance using OA-ML-APA again catches up with OLS. At intermediate time scales, increasing the projection order $P$ improves the convergence rate of OA-ML-APA, bringing it closer to the offline performance.  

\begin{figure}[h]
\centering
        \includegraphics[width=0.45\textwidth]{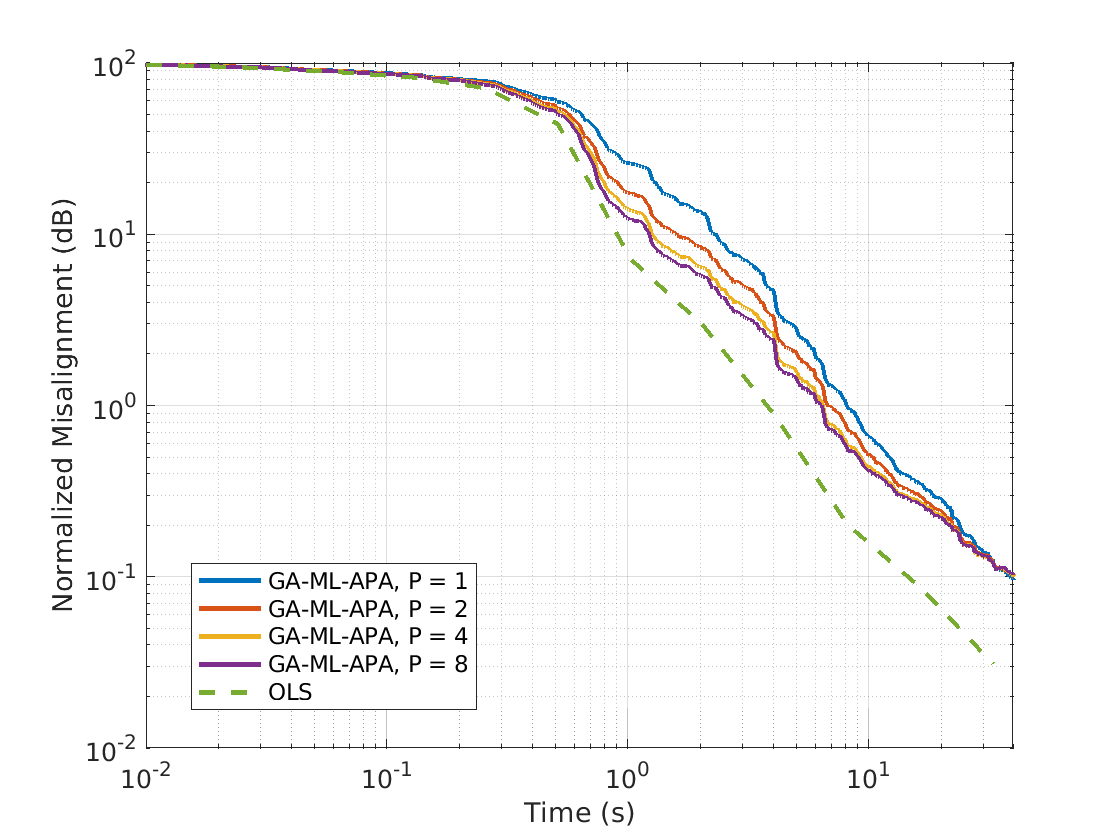}
         \caption{Comparing performace of OA-ML-APA and OLS when the input is recorded speech samples, for $L=512$ and for $P=1,2,4,8$}
         \label{fig:oaml_vs_ols_with_speech}
\end{figure}

In Figure~\ref{fig:oaml_vs_ols_with_speech}, we repeat the experiment using Librispeech audio samples as input. Audio traces have strong correlation and a much more complex statistical structure.  In this case, there is an apparent gap between the asymptotic rate of convergence of OLS and that of the online methods. It is still the case that OA-ML-APA brings down the misalignment quickly, and that the intermediate-term performance is improved by using larger values of $P$.

\subsection{Block updates versus incremental updates}\label{sim:block_vs_incremental}

In this paper, we used maximum likelihood arguments to derive two different APA updates rules: ML-APA when updating the filter estimate once every $P$ samples, and IML-APA when updating the filter estimate after every received sample. Our current theoretical analysis applies only to ML-APA, but empirically we have observed that IML-APA performs very similarly. In this subsection, we compare the performance of ML-APA and IML-APA. We also compare with two other methods of adapting ML-APA to give incremental updates; the results illustrate that the maximum likelihood approach used to derive IML-APA is a useful technique.

Recall the ML-APA update formula
$$
\wv_{t+P}=\wv_t+ X_t({1\over c_t}I_P+  X_t^TX_t)^{-1}(\yv_t-X_t^T\wv_t).
$$
A naive way to adapt this formula to update incrementally would be to simply update the weights after every sample, using the errors observed over the last $P$ samples, as is done in standard APA.  However, this approach turns out to be unstable. Intuitively, the method can be stabilized by reducing the update magnitude by a factor of $1/P$, obtaining the rule
$$
\wv_{t+1}=\wv_t+ {1\over P} X_t({1\over c_t}I_P+  X_t^TX_t)^{-1}(\yv_t-X_t^T\wv_t).
$$
Empirically, this stabilized approach performs similarly to ML-APA and IML-APA. However, both ML-APA and IML-APA are less complex to implement, since ML-APA only requires an update once every $P$ samples, and IML-APA requires matrix inversion of size $P-1$ rather than of size $P$.

\begin{figure}[h]
\centering
        \includegraphics[width=0.45\textwidth]{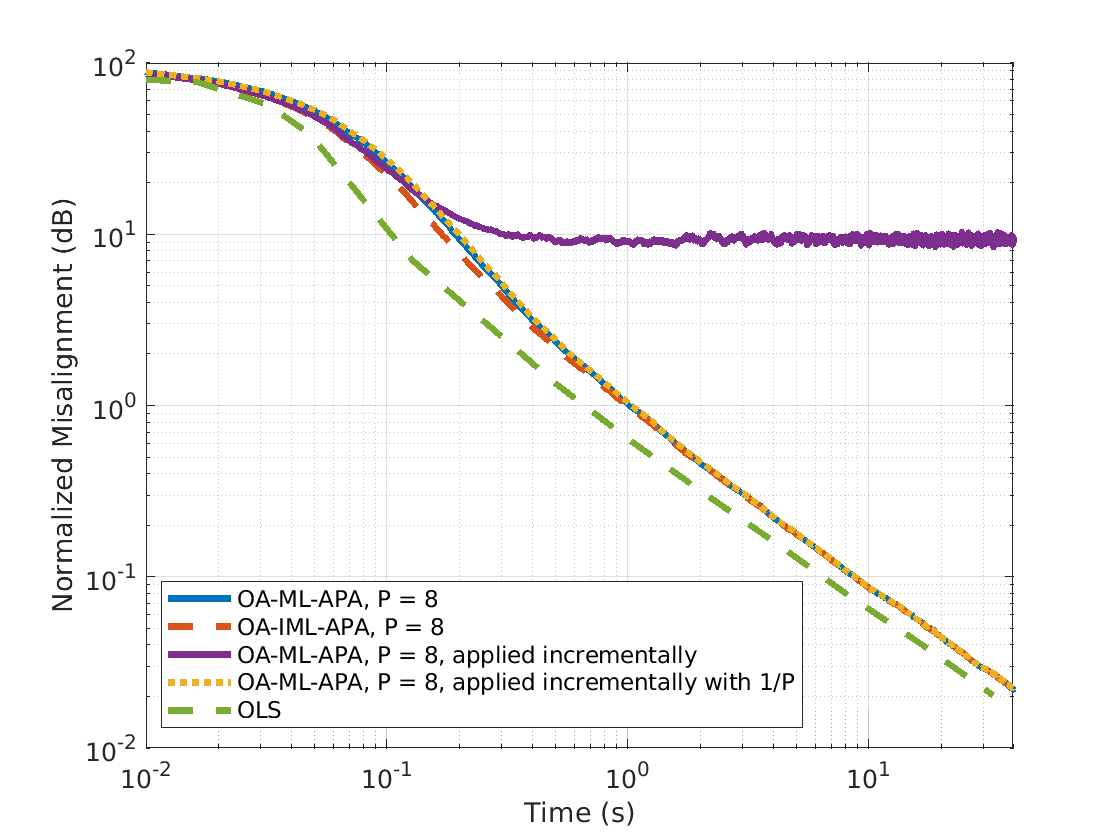}
         \caption{Comparing performace of OA-ML-APA, OA-IML-APA, and two incremental variants of OA-ML-APA when the input is an autoregressive process, for $L=512$ and for $P=8$}
         \label{fig:oaml_vs_oaiml_vs_2variants_vs_ols_with_ar}
\end{figure}

\begin{figure}[h]
\centering
        \includegraphics[width=0.45\textwidth]{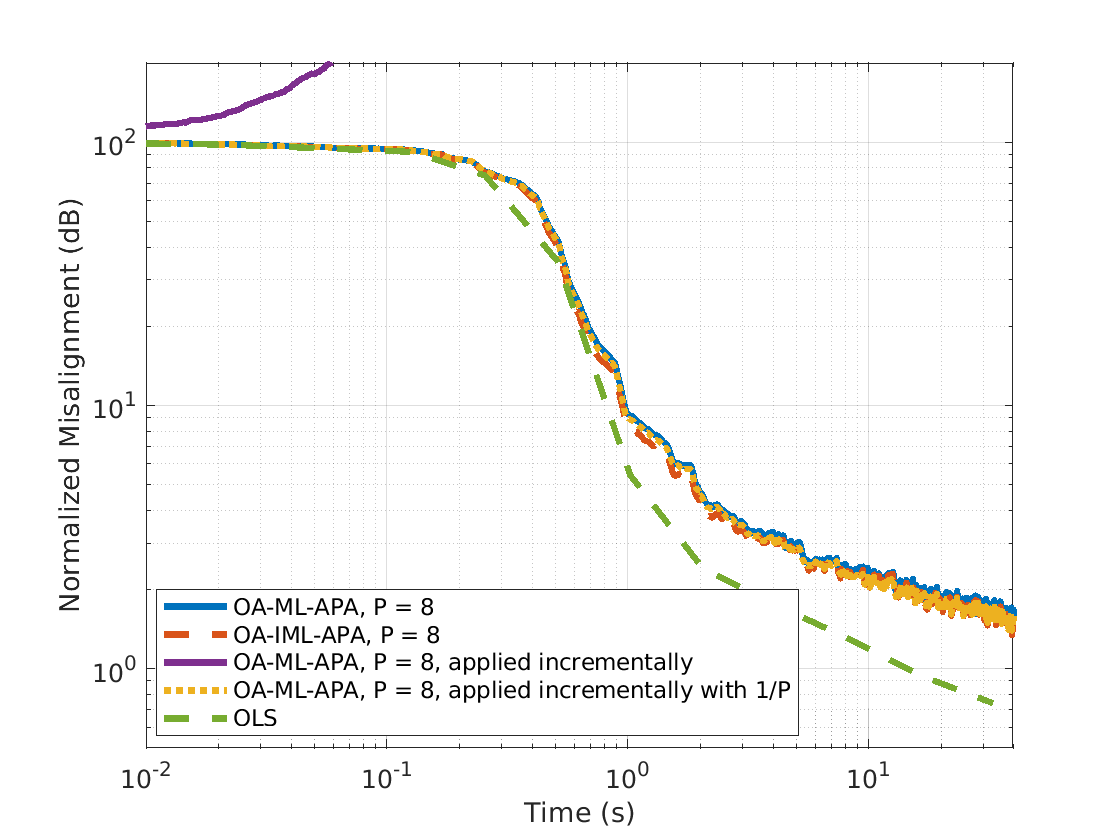}
         \caption{Comparing performace of OA-ML-APA, OA-I and OLS when the input is recorded speech samples, for $L=512$ and for $P=8$}
         \label{fig:oaml_vs_oaiml_vs_2variants_vs_ols_with_speech}
\end{figure}

Figures~\ref{fig:oaml_vs_oaiml_vs_2variants_vs_ols_with_ar} and \ref{fig:oaml_vs_oaiml_vs_2variants_vs_ols_with_speech} compare the performance of four methods, with autoregressive and speech inputs, respectively. The OLS performance is again included for reference. The projection order is set to $P=8$, with all other conditions as in Section~\ref{sim:corr_input}. 
As the figures show, the two ML based methods, OA-ML-APA and OA-IML-APA, perform very similarly to each other, despite OA-ML-APA being block-based and OA-IML-APA updating every sample. The incremental approach OA-IML-APA converges slightly faster than OA-ML-APA in the intermediate time scales. The figures also show that naive method of applying the OA-ML-APA incrementally does not perform well. With the autoregressive input, the naive method performs similarly to a fixed-step size APA algorithm, in that the misalignment saturates and does not improve asymptotically. Worse, with the speech input, the filter fails to converge.  After heuristically stabilizing this approach by dividing the step size by $P$, the stabilized method performs almost identically to OA-ML-APA, albeit while requiring $P$ times the complexity to implement.  Thus, OA-ML-APA appears to be the best choice of block update, while OA-IML-APA is the best approach to do incremental updates.The  stabilized update formula is essentially the same as that of the jointly optimized APA approach defined in \cite{ciochinua2015optimized}, if the misalignment estimate were provided by an oracle.

\subsection{Estimated misalignment versus oracle methods}\label{sim:de_vs_oa}

Up to this point, we have presented results for ML-APA and IML-APA that rely on an oracle to provide the current filter misalignment in each update cycle, allowing a precise calculation of the misalignment to noise ratio. In practice, the misalignment is not known, and needs to be estimated from available information. In this section, we describe one such method, and show empirically its usefulness.
The method we describe is effective in a  basic scenarios with known noise variance $\sigma_z^2$ and correctly handles discrete channel impulse changes.
Our point is not to claim that this method will work without modification in all scenarios found in real-world application, such as dynamic noise, double talk, and more complex channel dynamics.
Rather, we want to demonstrate that perfect knowledge of the misalignment is not necessary for these ML-based methods to work in practice.

 Our estimate of $m_t$ is obtained by combining two estimates with complementary properties.
 Defining $e_t = y_t - \wv_t^T \xv_t = (\wv^* - \wv^t)^T \xv_t + z_t$, assuming low correlation in $\xv_t$ and independence of $\xv_t$ and $z_t$, we have for fixed $\wv_t$ that $\E[ e_t^2] \approx E[x_t^2] \| \wv^* - \wv_t \|^2 + \sigma_z^2$.  We can thus  estimate $\hat{m}_t^{(1)} = (\E[e_t^2] - \sigma_z^2 )/( L E[x_t^2] )$, where ``local" estimates of  $\E[e_t^2]$ and $\E[x_t^2]$ can be obtained by exponential averaging. This estimate $\hat{m}_t^{(1)}$ can be trusted when $\E[e_t^2] \gg\sigma_z^2$. Thus it is most useful at initialization or after a sudden channel change, when the error dominates the noise level. This estimate fails to give good estimates of misalignment after convergence when the error is dominated by noise.

The second  estimate uses the delay-and-extrapolate method \cite{hansler2005acoustic} . Here, the input signal is artificially delayed  by $D$ samples so that by causality, the first $D$ coefficients of $\wv^*$ are known to be zero. Assuming uniform  error distribution over all   coefficients, we estimate 
\[
\hat{m}_t^{(2)} = {1\over D} \sum_{i=0}^{D-1} (w_{t,i}-w_i^*)^2 = {1\over D} \sum_{i=0}^{D-1} w_{t,i}^2.
\] This estimate is complementary to the other estimate, in that it is accurate when $\wv_t$ has converged to a nearly unbiased estimate of $\wv^*$, but  it need not be accurate initially or after a sudden channel change.

Combining these two estimates as indicated below, we obtain  delay-extrapolate ML-APA (DE-ML-APA). The method assumes that the first $D$ coefficients of $\wv^*$ are equal to zero. 
\begin{algorithm}
\caption{DE-ML-APA}\label{alg:estimate-misalignment}
\begin{algorithmic}
\Require $(\xv_1,\ldots,\xv_T)$,  $(y_1,\ldots,y_T)$, $D$, $L$ and $\sigma_z^2$ and $\epsilon$
\Ensure $\wv_t$
\State{Initialize $\wv\gets {\bf 0}_L$, $\hat{\sigma}_x^2 \gets 0$, $\hat{\sigma}_e^2 \gets 0$}
\For{$t \gets 1$ to $T$,  step = $P$}
    \State{ $X\gets [\xv_t,\ldots,\xv_{t-P+1}]$}
    \State{ $\yv \gets [y_t,\ldots,y_{t-P+1}]^T$}
    \State{ $\ev\gets \yv-X^T\wv$}
 \State{ $\hat{\sigma}_x^2 \gets (1-P/L) \hat{\sigma}_x^2 + {1\over L} \sum_{i=1}^P X_{1,i}^2$ }
 \State{ $\hat{\sigma}_e^2 \gets (1-P/L) \hat{\sigma}_e^2 +{1\over L} \sum_{i=1}^P e_i^2$}
 \State{ If $\hat{\sigma}_e^2 > 2\sigma_z^2$, $\hat{m}^{(1)} \gets \frac{ \hat{\sigma}_e^2 - \sigma_z^2}{(\hat{\sigma}_x^2 + \epsilon) L}$, else $\hat{m}^{(1)} \gets \epsilon$ }
 \State{ $\hat{m}^{(2)} \gets {1\over D}\sum_{i=0}^{D-1} w_{i}^2$}
 \State{ $\hat{m} \gets \max \{ \hat{m}^{(1)} , \hat{m}^{(2)} \} $}
  \State{ $c\gets \hat{m}/\sigma_z^2 $ }
  \State{ $\wv\gets \wv+ X({1\over c}I_P+  X^TX)^{-1} \ev $}
\EndFor
\end{algorithmic}
\end{algorithm}

\begin{figure}[h]
\centering
        \includegraphics[width=0.45\textwidth]{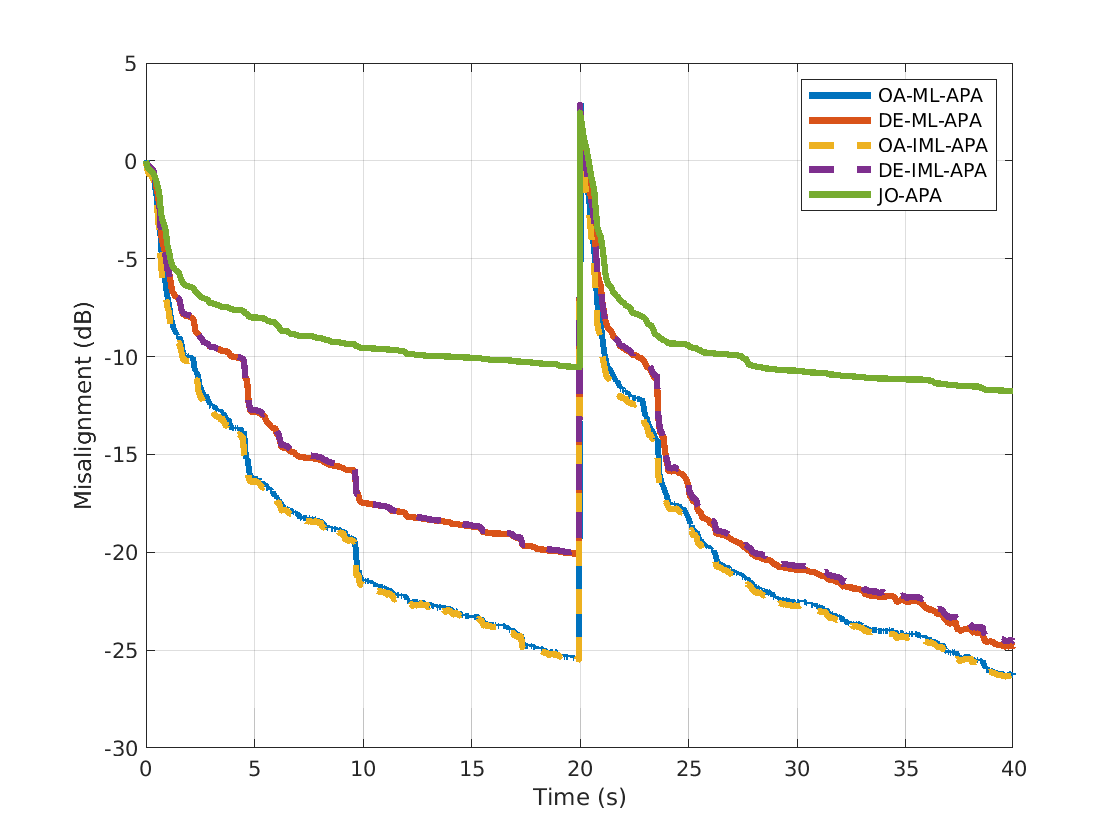}
         \caption{Comparing performace of  two oracle aided algorithms (OA-ML-APA and OA-IML-APA) with three non-oracle methods (DE-ML-APA, DE-IML-APA, and JO-APA, for speech inputs. A sudden channel change is introduced after 20 s. }
         \label{fig:oaml_vs_deml_vs_iml_vs_joapa_with_speech}
\end{figure}

Fig.~\ref{fig:oaml_vs_deml_vs_iml_vs_joapa_with_speech} shows an experiment using the same system parameters as for Fig.~\ref{fig:oaml_vs_oaiml_vs_2variants_vs_ols_with_speech} except that $P = 4$ , the simulation time is extended to 40 seconds, and and as in \cite{ciochinua2015optimized} the channel impulse response is shifted by 12 samples after 20s  to simulate a sudden change in an acoustic echo environment. 

We applied the  oracle-aided methods studied so far, OA-ML-APA and OA-IML-APA, as well as their non-oracle counterparts, DE-ML-APA and DE-IML-APA. We also tested the jointly-optimized APA (JO-APA) algorithm defined in \cite{ciochinua2015optimized}.  In each case, we averaged the 
observed misalignment over 10 independent trials, and plotted the result against time. The algorithms are all qualitatively similar in quickly bringing down the misalignment initially and after the sudden channel change, and they all continue to reduce misalignment further after initial convergence.  The DE-ML-APA algorithms and DE-IML-APA algorithms perform nearly identically to each other. The JO-APA algorithm has fast initial convergence, but its misalignment improves much more slowly than for the other algorithms after initial convergence.

\begin{figure}[h]
\centering
        \includegraphics[width=0.45\textwidth]{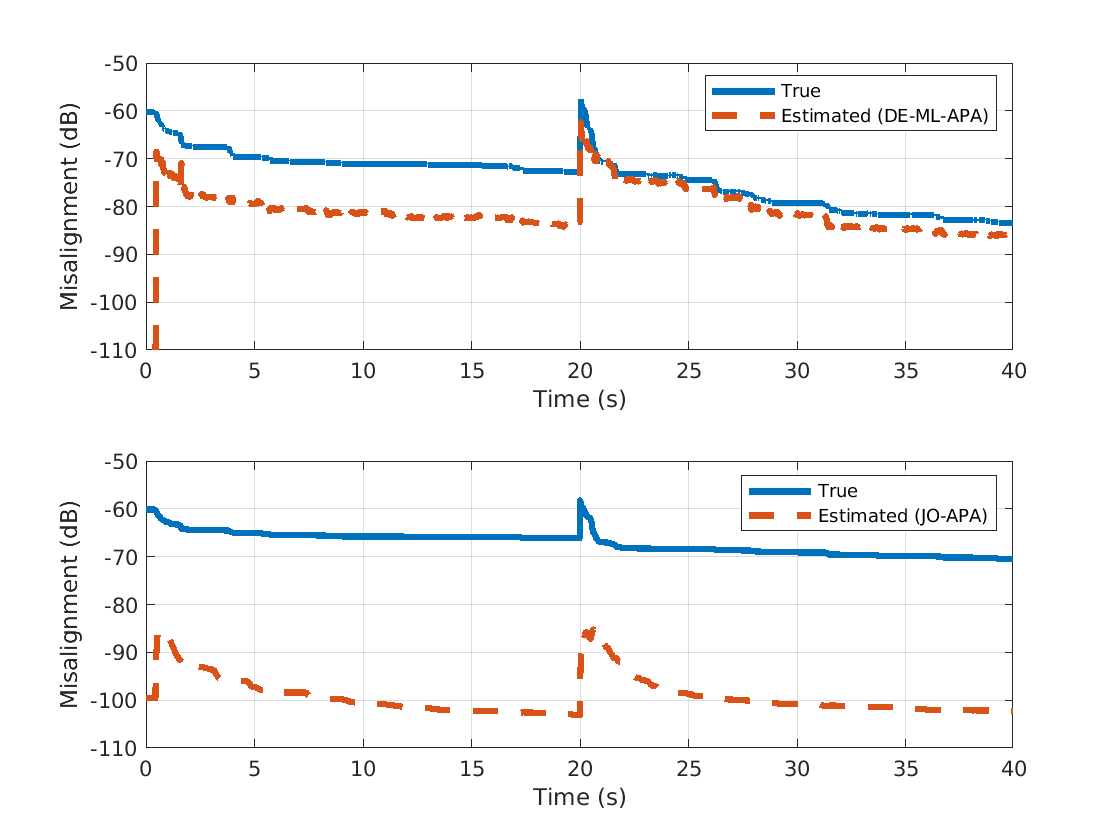}
         \caption{For the experiments reported in Figure~\ref{fig:oaml_vs_deml_vs_iml_vs_joapa_with_speech}, the evolution of true misalignment and estimated misalignment with time, for (top) the DE-ML-APA algorithm and (bottom) the JO-APA algorithm}
         \label{fig:detail_oaml_vs_deml_vs_iml_vs_joapa_with_speech}
\end{figure}

Further analysis shows that the gap in performance between JO-APA and DE-ML-APA is mainly due to DE-ML-APA having a more accurate estimate of the misalignment in this setting. In the top half of Figure~\ref{fig:detail_oaml_vs_deml_vs_iml_vs_joapa_with_speech}, we show the actual evolution of misalignment under the DE-ML-APA algorithm together with the evolution of the estimated misalignment available to the algorithm at each step. Similarly, the bottom of the figure shows the actual evolution of the misalignment uder the JO-APA algorithm together with the estimated misalignment available to that algorithm in each step. While the true misalignment is underestimated in both cases, the misalignment estimate obtained with DE-ML-APA is much more accurate than that of JO-APA.

\section{Proof of the main results}\label{sec:proof}
Before presenting  the proof of Theorem \ref{thm:main}, we reproduce a lemma from
\cite{RudelsonVershinin2010} that we will need about the concentration of the singular values of i.i.d.~random matrices.

\begin{lemma}\label{lem:singvalues}\label{lemma:sigma}
Let the elements of an $m \times n$ matrix $A$, $m<n$, be drawn independently from $\Nc(0,1)$. Then, for any $h>0$,
\begin{align*}
\P\Big(&\sqrt{n}-\sqrt{m}-\sqrt{h}\leq  \sigma_{\min} (A) \nonumber\\
&\leq  \sigma_{\max}(A)\ \leq \sqrt{n}+\sqrt{m}+ \sqrt{h}\Big) 
\geq 1-  2 {\rm e}^{-\frac{h}{2}}.
\end{align*}
\end{lemma}

\begin{proof}[Proof of Theorem \ref{thm:main}]
Recall that $\wv_{t+P}=\wv_t+ X_t({1\over c_t}I_P+  X_t^TX_t)^{-1}(\yv_t-X_t^T\wv_t)$,
where we define $r_t=\|\wv_t-\wv^*\|^2$ and $c_t={  r_t\over L\sigma_z^2}$. 
Since $\yv_t=X_t^T\wv^*+\zv_t$, it follows that
\begin{align*}
\wv_{t+P}-\wv^*=&A_t (\wv_t-\wv^*)+X_t({ 1\over c_t }I_P+  X_t^TX_t)^{-1}\zv_t,
\end{align*}
where 
\begin{align}
A_t= I_L- X_t({ 1\over c_t}I_P+  X_t^TX_t)^{-1}X_t^T.\label{def:At}
\end{align}
Since, $\zv_t$ is zero-mean and independent of $X_t$, we have
\begin{align}
\E[\|\wv_{t+P}-\wv^*\|^2]=&\E[\|A_t (\wv_t-\wv^*)\|^2]\nonumber\\
&+\E[\|X_t({ 1\over c_t }I_P+  X_t^TX_t)^{-1}\zv_t\|^2].\label{eq:main-error-t}
\end{align}
For $t=1,2,\ldots$, define the (misalignment) error vector $\ev_t$ as
\[
\ev_t=\wv_{t}-\wv^*.
\]
Also, recall that earlier, we defined $r_t$ as $r_t=\|\ev_t\|^2$. Then, given the independence of $X_t$ and $\zv_t$,  from \eqref{eq:main-error-t} it follows that
\begin{align}
\E[\|\ev_{t+P}\|^2]=&\E[\ev_t^TA_t^2 \ev_t]\nonumber\\
&+\sigma_z^2\E[\trace(X_t({ 1\over c_t }I_P+  X_t^TX_t)^{-2}X_t^T)].\label{eq:main-error-t-simple}
\end{align}

The random matrix $X_t$ is rotationally invariant, i.e. $\tilde{X} = U X_t$ has the same distribution as $X_t$ for any fixed orthonormal $U\in\mathds{R}^{L\times L}$.  Therefore, it is easy to see that the distribution of $A_t^2$ is invariant to the transformation $U A_t^2  U^T$, and thus $\E[\ev_t^TA_t^2 \ev_t]$ depends on $\ev_t$ only through its norm $\sqrt{r_t}$. Denoting by
$\tilde{\ev}_m$ the $m$-th column of $\sqrt{r_t} I$, for each $m$, we have
$$
\E[\ev_t^TA_t^2\ev_t] = \E[\tilde{\ev}_m^TA_t^2 \tilde{\ev_t}_m]=r_t \E[\left(A_t^2 \right)_{mm}]. 
$$
Summing over $m$, it follows that $\E[\ev_t^TA_t^2 \ev_t] = \frac{r_t}{L} \E[ \trace (A_t^2 )]$.

It is convenient to define the normalized misalignment $a_t = r_t \sigma_x^2/\sigma_z^2$. In an echo cancellation setting, this is also known as the normalized misadjustment: the ratio of additional output power due to imperfect filter coefficients, divided by the output power obtained with perfect coefficients. In these terms, we have
\begin{align}
\E[a_{t+P}]\ =  &\frac{a_t}{L}  \E[ \trace (A_t^2 )] 
+\sigma_x^2\E[\trace(X_t({ 1\over c_t }I_P+  X_t^TX_t)^{-2}X_t^T)].\label{eq:main-error-t-simple-a}
\end{align}

Consider the singular value decomposition (SVD) of $X_t$ defined as $X_t=USV^T$, where $U\in\mathds{R}^{L\times P}$, $S\in\mathds{R}^{P\times P}$ and $V\in\mathds{R}^{P\times P}$.
Denoting the $P$ singular values as $S_1, \ldots, S_P$, we can compute the eigenvalues of the matrices in (\ref{eq:main-error-t-simple-a}).
The matrix $A_t$  has $L-P$ eigenvalues equal to 1, and the remaining $P$ eigenvalues equal to $1-\frac{S_i^2}{c_t^{-1} + S_i^2} = \frac{1}{1 + c_t S_i^2}$, while the matrix $X_t({ 1\over c_t }I_P+  X_t^TX_t)^{-2}X_t^T$ has $L-P$ eigenvalues equal to zero and the remaining $P$ eigenvalues equal to $ \frac{ c_t^2  S_i^2}{(1 + c_t S_i^2)^2}$. Thus
\begin{align}
\E[a_{t+P}] =  &\frac{a_t}{L} \left( L - P + \sum_{i=1}^P \E\left[ \frac{1}{(1 + c_t S_i^2)^2}\right]\right)   \nonumber \\
&+\sigma_x^2\sum_{i=1}^P \E\left[\frac{ c_t^2  S_i^2}{(1 + c_t S_i^2)^2} \right]. \nonumber \\
= & a_t(1 - \frac{P}{L}) + \frac{a_t}{L} \sum_{i=1}^P \E \left[ \frac{1}{(1 + a_t\tilde{ S}_i^2)^2} + 
 \frac{ a_t \tilde{ S}_i^2}{(1 + a_t  \tilde{ S}_i^2)^2} \right] \nonumber \\
=  & a_t \left( 1 - \frac{P}{L} + \frac{1}{L}  \sum_{i=1}^P \E\left[ \frac{1}{1 + a_t\tilde{ S}_i^2}\right]\right) \nonumber \\
= & a_t \left( 1 -  \frac{1}{L}  \sum_{i=1}^P \E\left[ \frac{a_t \tilde{S}_j^2 }{1 + a_t\tilde{ S}_i^2}\right] \right), \nonumber 
\end{align}
where we have defined normalized singular values $\tilde{S}_i = S_i/\sqrt{L\sigma_x^2}$, and used $c_t = r_t/(L\sigma_z^2) = a_t/(L\sigma_x^2)$.

To complete the proof, we will show that
$$
  \E\left[  \frac{a_t \tilde{S}_j^2 }{1 + a_t\tilde{ S}_i^2}\right] \geq   \left(1-\gamma_L\right) \frac{a_t}{1+a_t}
$$
for a constant $\gamma_L$ that goes to zero, as $L\to\infty$.

This can be established because the normalized singular values concentrate around unity for large $L$.
For convenience, define 
\[
\psi(S,a) = {a  S^2\over 1+a S^2}.
\]
Since for any $a_t>0$,  $\psi(S,a_t)$ is non-negative and increasing in $S>0$, for any $\tau \geq 0$, we have
\begin{align}
\E[ \psi(\tilde{S}_i,a_t) ]  = & E[ \psi(\tilde{S}_i,a_t) | S_i < \tau ] P[\tilde{S}_i < \tau]\\
&  + E[ \psi(\tilde{S}_i,a_t) | \tilde{S}_i \geq \tau ] P[\tilde{S}_i \geq \tau] \\
\geq & \psi(\tau,a_t) P[\tilde{ S}_i \geq \tau].
\end{align}
  Moreover, if $\tau \leq 1$, $\psi(\tau,a_t) \geq \psi(1,a_t)\tau^2$, so for any $ \tau \in(0,1)$, we have
$$
\E[ \psi(\tilde{S}_i,a_t) ]  \geq \psi(1,a_t) \tau^2 P[\tilde{ S}_i \geq \tau].
$$
On the other hand, from Lemma~\ref{lemma:sigma}, we know that
$$
P\left[\tilde{S}_i  \geq 1 - \sqrt{\frac{P}{L}} - \sqrt{\frac{h}{L}} \right] \geq 1 - 2 e^{-h/2}
$$
for any $0\leq h \leq \sqrt{L}-\sqrt{P}$. Thus taking $\tau = \left(1 - \sqrt{\frac{P}{L}} - \sqrt{\frac{h}{L}}\right)$, we have
$$
\E[ \psi(\tilde{S}_i,a_t) ] \geq \psi(1,a_t)  \left(1 - \sqrt{\frac{P}{L}} - \sqrt{\frac{h}{L}}\right)^2\left(1-2e^{-h/2}\right)
$$
Choosing $h = \log L$, assuming $\log L \leq \sqrt{L}-\sqrt{P}$,  we obtain the desired bound with
$$
\gamma_L = 1 - \left(1 - \sqrt{\frac{P}{L}} - \sqrt{\frac{\log L}{L}}\right)^2\left(1-\frac{2}{\sqrt{L}} \right).
$$
For large $L$, 
$$
\gamma_L \approx 2 \frac{\sqrt{P}+1 + \sqrt{\log{L}}}{\sqrt{L}}
$$
establishing that $\gamma_L = O(\sqrt{\log L\,/\,L})$.
\end{proof}

\begin{proof}[Proof of Theorem \ref{thm:offline}]
The arguments in the proof of Theorem~\ref{thm:main} up to equation (\ref{eq:main-error-t-simple-a}) apply here too, except that we need to  substitute a generic regularization parameter $\delta$ and consider that $X_s$ now includes the past $s$ samples instead of the past $P$ samples. Also, in the case that $\delta =0$ matrix inverses are replaced by pseudo-inverses.  Instead of (\ref{eq:main-error-t-simple-a}), we have
\begin{align}
\E[a^{(\delta)}_s]\ =  &\frac{a_0}{L}  \E[ \trace (A_t^2 )] 
+\sigma_x^2\E[\trace(X_t(\delta I_s+  X_t^TX_t)^{-2}X_t^T)],\label{eq:main-error-t-simple-a-offline}
\end{align}
where $A_t$ is defined in \eqref{def:At}. 
When $s \leq L$,  $A_t$ has $L-s$ eigenvalues equal to 1, and $s$ eigenvalues equal to $\frac{\delta}{\delta + S_i^2}$, 
while the matrix $X_t(\delta I_s+  X_t^TX_t)^{-2}X_t^T$ has $L-s$ eigenvalues equal to zero and the remaining $s$ eigenvalues equal to $ \frac{   S_i^2}{(\delta +  S_i^2)^2}$. Thus
\begin{align}
\E[a^{(\delta)}_s] =  &\frac{a_0}{L} \left( L - s + \sum_{i=1}^s \E\left[ \frac{\delta^2}{ ( \delta + S_i^2)^2}\right]\right)   \nonumber \\
{} & + \sigma_x^2
 \sum_{i=1}^s 
 \E[  \frac{  S_i^2} {(\delta +  S_i^2)^2} ]
 \nonumber \\
= & a_0(1 - \frac{s}{L}) + \sigma_x^2  \sum_{i=1}^s \E \left[ \frac{ c_0 \delta^2 + S_i^2 }{(\delta +   S_i^2)^2}  \right], \nonumber 
\end{align}
where, as before, we have defined $c_0 = a_0/(L\sigma_x^2)$.   By  examing the first derivative, it may be easily verified that for $S^2>0$ and $c>0$, the function $(c \delta^2 + S^2)/(\delta + S^2)^2$ has a unique minimum on 
$\delta \geq 0$, achieved at $\delta = 1/c$, and equal to $1/(c^{-1}+S^2)$. If $S^2=0$, the function is constant and equal to $c$; so the same bound holds in that case.  Since $\delta = c_0^{-1}$ is the minimizer regardess of the value of the singular value $S_i$, we have $\E[a^{(\delta)}_s] \geq \E[a^{(c_0^{-1})}_s]$. Further,
\begin{align}
\E[a^{(c_0^{-1})}_s ]&  \geq  a_0(1 - \frac{s}{L}) + \sigma_x^2  \sum_{i=1}^s \E \left[ \frac{ 1 }{c_0^{-1} +   S_i^2}  \right] \nonumber \\
& \geq  a_0(1 - \frac{s}{L}) +  \sigma_x^2  s  \E \left[ \frac{ 1 }{c_0^{-1} +   S_I^2}  \right] \nonumber \\
& \geq  a_0(1 - \frac{s}{L}) +  \sigma_x^2  s  \frac{ 1 }{c_0^{-1} +   L\sigma_x^2}  \label{eq:jensen2}\\
& \geq  a_0(1 - \frac{s}{L}) +    \frac{s}{L}  \frac{ a_0 }{a_0+1} \nonumber \\
& \geq a_0\left( 1 - \frac{s}{L}\ \frac{a_0}{1+a_0} \right).
\end{align}
Here, $S_I$ denotes a singular value chosen uniformly at random from among the $s$ singular values, with $\E[ S_I^2] = L\sigma_x^2$, and (\ref{eq:jensen2}) is obtained by the Jensen's inequality.

For the case $s \geq L$, $A_s$ has $L$ eigenvalues equal to $\frac{\delta}{\delta + S_i^2}$ and 
$X_t(\delta I_s+  X_t^TX_t)^{-2}X_t^T$ has $s-L$ eigenvalues equal to zero and the remaining $L$ eigenvalues equal to $ \frac{   S_i^2}{(\delta +  S_i^2)^2}$. Thus
\begin{align}
\E[a^{(\delta)}_s] =  &\frac{a_0}{L}  \sum_{i=1}^L \E\left[ \frac{\delta^2}{ ( \delta + S_i^2)^2}\right]   \nonumber \\
{} & + \sigma_x^2
 \sum_{i=1}^L 
 \E[  \frac{  S_i^2} {(\delta +  S_i^2)^2} ]
 \nonumber \\
= &  \sigma_x^2  \sum_{i=1}^L \E \left[ \frac{ c_0 \delta^2 + S_i^2 }{(\delta +   S_i^2)^2}  \right]. \nonumber 
\end{align}
Minimizing each term over $\delta$ as before, we obtain
\begin{align}
\E[ a^{(c_0^{-1})}_s ]&  \geq  \sigma_x^2  \sum_{i=1}^L \E \left[ \frac{ 1 }{c_0^{-1} +   S_i^2}  \right] \nonumber \\
& \geq   L \sigma_x^2    \E \left[ \frac{ 1 }{c_0^{-1} +   S_I^2}  \right] \nonumber \\
& \geq  L \sigma_x^2    \frac{ 1 }{c_0^{-1} +  s \sigma_x^2}  \label{eq:jensen3}\\
& \geq    \frac{L a_0}{L + s a_0}, \nonumber 
\end{align}
where we again use the Jensen's inequality, and in this case have $\E[ S_I^2] = s \sigma_x^2$.

\end{proof}



\section{Conclusions}\label{sec:conclude}

We have introduced a maximum-likelihood based parameter control scheme for APA (ML-APA), and proved that when an oracle provides the misalignment to noise ratio (MNR), and the input is i.i.d., the algorithm is near-optimal at any finite time and asymptotically achieves the same performance as offline least squares. Empirically, we show that the oracle-driven scheme also performs close to offline least squares for realistic speech inputs. We show that the algorithm also works well when an estimate of the MNR is substituted for the oracle. In a scenario with speech input and a dynamic impulse response, the estimate-driven algorithm performs nearly as well as the oracle-based version, achieving fast convergence and continuous steady state improvement. 

The combination of analysis and simulation show that ML-APA and its incremental counterpart IML-APA provide  an excellent framework for high-performance adaptive filtering. To use ML-APA in practice, robust methods for estimating MNR are needed;  this is an area of current study for us. Extending the analysis to correlated and non-stationary sources is also an area of study that may yield further insights.

\appendices

\section{Mean behavior of ML-APAL}\label{app:b}

\begin{lemma}\label{lemma:app-b}
Consider the sequence $a_{t+1} = f(a_t)$ with initial condition $a_1>0$ and 
$$
f(a) = \left(1-b\frac{a}{1+a}\right) a
$$
for some $0< b < 1$. Then for any $\alpha > 0$, there exists $T$ such that
$$
(1-\alpha) (bt)^{-1} \leq a_t \leq (1+\alpha) (bt)^{-1}
$$
for $t\geq T$.
\end{lemma}
\begin{proof}
For any $a > 0$, we have $0 < f(a) < a$, so $a_t$ is a decreasing sequence with limit point $f(0) = 0$. We will show that for any initial condition, eventually $a_t \sim (bt)^{-1}$.

The function can be expressed $f(a) = (1-b)a + b\frac{a}{1+a}$, hence $f'(a) = 1-b + \frac{b}{(1+a)^2}$ and we see that $f$ is monotonically increasing for $a\geq 0$.

{\em Lower bound.}   We will show that the sequence $\gamma_t := a_t^{-1}-bt$ is decreasing. Then for the lower bound, given $\alpha > 0$, we set $T \geq \gamma_1 (1-\alpha)/(\alpha b)$.  For $t\geq T$, we have
$$
b t a_t = \frac{bt}{bt + \gamma_t} \geq \frac{bt}{bt + \gamma_1} \geq \frac{bT}{bT+\gamma_1} \geq 1-\alpha.
$$

To show that $\gamma_t$ is decreasing, we use  $f(a) = (1-b)a + ba/(1+a)$, such that
\begin{eqnarray*}
a_{t+1} & = & f\left( \frac{1}{bt+\gamma_t}\right)  \\
& = & \frac{1-b}{bt+\gamma_t} + \frac{b}{bt+\gamma_t+1} \\
& = & \frac{bt + \gamma_t + (1-b)}{(bt+\gamma_t)(bt + \gamma_t+1)}
\end{eqnarray*}
Further algebraic manipulation then shows that
$$
\gamma_{t+1} = a_{t+1}^{-1} - b(t+1) = \gamma_t - \frac{b(1-b)}{bt + \gamma_t+(1-b)} < \gamma_t.
$$

{\em Upper bound.}
Given $a_t$, we define the related sequence $\alpha_t = b t a_t - 1$, so that $a_t = (1+\alpha_t)(bt)^{-1}$.
For the upper bound, for any initial condition $a_1$ we construct a sequence $\beta_t$ with $\beta_t \geq \alpha_t$  and with $\beta_t$ converging to zero.

First we establish a bound on $\alpha_{t+1}$ given a bound  $a_t \leq (1+\beta_t)(bt)^{-1}$. 
We have
\begin{eqnarray*}
\alpha_{t+1} & = & b (t+1) a_{t+1} - 1  = b(t+1)f(a_t) - 1\\
& \leq & b(t+1) f\left( (1+\beta_t)(bt)^{-1}\right) - 1 \\
& \leq &  (1+\beta_t)\left[ 1 - b \frac{1+\beta_t}{bt + 1 + \beta_t}\right] \frac{t+1}{t} -1\\
& \leq &  (1+\beta_t)\left[ \frac{bt + (1-b)(1+\beta_t)}{bt + 1 + \beta_t}\right] \frac{t+1}{t} -1\\
& \leq &  (1+\beta_t)\left[ 1 - \frac{bt \beta_t - (1-b)(1+\beta_t)}{t(bt + 1 + \beta_t)}\right] -1\\
& \leq & \beta_t -  \frac{(1+\beta_t)(bt \beta_t - (1-b)(1+\beta_t))}{t(bt + 1 + \beta_t)} \\
& \leq & \beta_t\left[ 1 - b \frac{1 + \beta_t}{(bt+1+\beta_t)}\left(1 - \frac{(1-b)(1+\beta_t)}{b t \beta_t}\right)\right]
\end{eqnarray*}
One consequence of this expression is that if we want our bound on $\alpha_{t+1}$ to be less than our bound $\alpha_t \leq \beta_t$, we need $(1-b)(1+\beta_t) < bt \beta_t$, or equivalently $\beta_t > (1-b)\left( bt - (1-b)\right)^{-1}$. Thus, we will design $\beta_t$ for $t > b^{-1}$ to meet the sufficient condition that $\beta_t \geq (bt-1)^{-1}$.

Suppose that for some $t > b^{-1}$, $\beta_t \geq \alpha_t$ and $\beta_t \geq (bt-1)^{-1}$. Then $\beta_t/(1+\beta_t) \geq (bt)^{-1}$ and $(1+\beta_t)/(bt+1+\beta_t) \geq (bt)^{-1}$. Our bound on $\alpha_{t+1}$ then becomes
\begin{eqnarray}
\alpha_{t+1} 
& \leq & \beta_t\left[ 1 - b \frac{1}{bt}\left(1 - (1-b) \right)\right] \nonumber \\
& \leq & \beta_t \left(1-\frac{b}{t}\right) \label{eqn:kicker}
\end{eqnarray}

To complete our definition of the sequence $\beta_t$, first suppose that $\a_t \leq 1/(bt-1)$ for all $t>b^{-1}$. In this case, we can take $\beta_t = b t (bt-1)^{-1} - 1 = (bt-1)^{-1}$ for all $t > b^{-1}$. Then  $\beta_t \geq \alpha_t$ and $\beta_t \rightarrow 0$ as desired.

Otherwise, let $T$ be the first $T > b^{-1}$ with $a_T > 1/(bT-1)$. We define $\beta_T = \alpha_T$, and for $t > T$, recursively define $\beta_{t+1} = \max\left\{ \beta_t\left(1-\frac{b}{t}\right)\, , \, \frac{1}{bt-1} \right\}$.
By virtue of (\ref{eqn:kicker}), we have $\beta_t \geq \alpha_t$ for all $t \geq T$.

To see that $\beta_t \rightarrow 0$, consider the related sequence defined by $\beta^{*}_{t+1} = \beta^{*}_t(1-b/t)$.
Taking logs, $\log(\beta^{*}_{t+1}) = \log(\beta^{*}_{t}) + \log(1-b/t) \approx \log(\beta^{*}_{t}) - b/t $ for sufficiently large $t$. Since the sum of $1/t$ is infinite, $\beta^{*}_t$ decreases without bound.

For any $\alpha > 0$, recursively define 
$$\beta^{(\alpha)}_{t+1} = \max\left\{ \, \beta^{(\alpha)}_t\left(1-\frac{b}{t}\right)\, , \, \frac{1}{bt-1} \, , \, \alpha \,\right\}.$$  Clearly $\beta^{(\alpha)}_t \geq  \beta_t$, and $\beta^{(\alpha)}_t \to \alpha$ as $t\to\infty$.  Since this is true for any $\alpha > 0$, we have $\beta_t \rightarrow 0$.

\end{proof}

\bibliographystyle{unsrt}
\bibliography{myrefs}
\end{document}

%% file: defns.tex
\usepackage{xspace}
\usepackage{bbm}
\newcommand{\Nc}{\mathcal{N}}

\newcommand{\Xc}{\mathcal{X}}
\newcommand{\Yc}{\mathcal{Y}}

\newcommand{\ev}{{\bf e}}

\newcommand{\xv}{{\bf x}}
\newcommand{\yv}{{\bf y}}
\newcommand{\zv}{{\bf z}}

\newcommand{\vv}{{\bf v}}

\newcommand{\wv}{{\bf w}}

\def\a{\alpha}

\DeclareMathOperator\E{E}
\let\P\relax
\DeclareMathOperator\P{P}

\def\textiid{i.i.d.\@\xspace}
\newcommand\iid{\ifmmode\text{ i.i.d. } \else \textiid \fi}



%% file: notations.tex
\usepackage{subfig}
\newcommand{\trace}{{\mathbf{tr}}}

\DeclareFontFamily{U}{mathx}{\hyphenchar\font45}
\DeclareFontShape{U}{mathx}{m}{n}{
      <5> <6> <7> <8> <9> <10>
      <10.95> <12> <14.4> <17.28> <20.74> <24.88>
      mathx10
      }{}
\DeclareSymbolFont{mathx}{U}{mathx}{m}{n}
\DeclareMathAccent{\widecheck}{\mathalpha}{mathx}{"71}

\usepackage{bm}

